\documentclass [oneside,11pt,onecolumn,draftclsnofoot]{IEEEtran}

\usepackage{cite}

\ifCLASSINFOpdf
   \usepackage[pdftex]{graphicx}
   \graphicspath{{../pdf/}{../jpeg/}}
   \DeclareGraphicsExtensions{.pdf,.jpeg,.png}
\else
   \usepackage[dvips]{graphicx}
   \graphicspath{{../eps/}}
   \DeclareGraphicsExtensions{.eps}
\fi
\usepackage{float}

\usepackage[cmex10]{amsmath}
\usepackage{array}

\usepackage{mdwmath}
\usepackage{mdwtab}

\usepackage{multirow}
\usepackage {psfrag} 
\usepackage{eqparbox}
\begin{document}
%


\title{Joint Beamforming and Feature Detection for Enhanced Visualization of Spinal Bone Surfaces in Ultrasound Images }

\author{Saeed~Mehdizadeh,
        Sebastien~Muller,        Gabriel Kiss, Tonni~F.~Johansen,
        and~Sverre~Holm
\thanks{S. Mehdizadeh, was with the Department of Circulation and Medical Imaging, Norwegian University of Science and Technology, Trondheim, Norway and is now with Cowi AS, Oslo.}
\thanks{S. Muller is with the Department of Medical Technology, SINTEF, Norway.}
\thanks{G. Kiss is with with the Department of Circulation and Medical Imaging, NTNU, and  St. Olavs Hospital, Trondheim, Norway.}
\thanks { Tonni F. Johansen in with the Department of Circulation and Medical Imaging, NTNU, and SINTEF ICT Trondheim, Norway.}
\thanks{S. Holm is with the Department of Informatics, University of Oslo, Norway.}}

\maketitle

\begin{abstract}
We propose a framework for extracting the bone surface from B-mode images employing the eigen-space minimum variance (ESMV) beamformer and a ridge detection method. We show that an ESMV beamformer with a rank-1 signal subspace can preserve the bone anatomy and enhance the edges, despite an image which is less visually appealing due to some speckle pattern distortion. The beamformed images are post-processed using the phase symmetry (PS) technique. We validate this framework by registering the ultrasound images of a vertebra (in a water bath) against the   corresponding Computed Tomography (CT) dataset.  The results show a bone localization error in the same order of  magnitude as the standard delay-and-sum (DAS) technique, but  with  approximately  20$\%$   smaller standard deviation (STD) of the  image intensity distribution around the bone surface. This indicates a sharper bone surface detection. Further, the  noise  level inside  the bone shadow  is reduced  by 60$\%$. In  in-vivo  experiments, this framework is used  for imaging the spinal anatomy. We show that PS images obtained from this beamformer setup have sharper bone boundaries in comparison with the standard DAS ones, and  they are reasonably well separated from the surrounding soft tissue.\\
\end{abstract}


%

\section{Introduction}
\label{intro}
The imaging of bone structures is usually done using X-ray or Computed Tomography (CT). However, ionizing radiation, scanner time cost, and lack of portability are the limitations of these modalities.~Ultrasound  may address these issues  in many  applications.~Ultrasound imaging of bone tissue has been investigated in different clinical procedures, \textit{e.g.},~registration of bone in neurosurgeries and orthopedics~\cite{Amin:2003,Winter:2008}, guidance for diagnosis of skeletal fractures in emergency rooms~\cite{Stewart:2009}, and pain management interventions~\cite{Eichenberger:2004,Gofeld:2009}.~Particularly, in some applications, dealing with  the spine is of interest, \textit{e.g.} guidance for minimal invasive (MI) procedures in spinal surgery~\cite{Bonsanto:2005,Kolstad:2006}, and for administration of spinal anesthesia~\cite{Arzola:2007,Tran:2009,Narouze:2010, Grau:2005}.\\


Ultrasound imaging is a valuable modality for enhancing the safety of different puncture techniques in regional anesthesia~\cite{Grau:2005,Narouze:2010}. These procedures are mostly performed landmark based or blind~\cite{Narouze:2010}.~Ultrasound can facilitate these routines by visualizing the spinal anatomy, assisting to locate the puncture region before performing the injection procedure.~Further, ultrasound can be used as a real-time modality for needle trajectory control, or more effective placement of medication~\cite{Grau:2005}. However, in epidural injections, the spinal structures obstruct the ultrasound beams and makes the images noisy.\\
Another potential application of ultrasound is computer-assisted minimally invasive (MI) spinal surgery~\cite{Winter:2008}. The procedure may require the registration of the patient positioned for surgery with preoperatively acquired images. The restriction of minimal invasiveness, together with limited radiation exposure, point at ultrasound imaging as a good candidate~\cite{Winter:2008}. The other important procedure in MI spine surgery is the accurate localization of the target vertebra. Conventionally, localizing a vertebral level is performed by manual palpation  and   direct fluoroscopy. Thus, surgeons identify a specific anatomical landmark such as the sacrum, and, then, start counting under fluoroscopic control up to the targeted vertebral level~\cite{Otake:2012}. This approach exposes the patient to an undesirable level of radiation, and is prone to counting errors due to the similar appearance of vertebrae in projection images~\cite{Otake:2012}. Alternatively, ultrasound can improve  patient safety and decrease the risk of wrong level surgery~\cite{Otake:2012}.\\
In general, bone imaging using conventional ultrasound techniques is prone to higher level of artifacts in comparison with soft tissue imaging~\cite{Mauldin:2012, Tran:2009}. In the case of  the spine, images are filled with acoustical noise, and artifacts that can impede visualization of important features, and  also make it hard to detect  and segment the bone structure. Image enhancement, where bone structures stand out more distinctly from surrounding soft tissue, helps to isolate the bone surface out of the B-mode ultrasound.\\
To automate segmentation of the bone structure, image intensity or gradient-based methods are common~\cite{Kowal:2007}, but results are sensitive to the  parameters of image acquisition, {e.g.}~frequency and dynamic range. Pattern recognition or statistical shape models provide more robust results but require learning sets, and fail to identify traumatic cases as the pattern searched for is disrupted~\cite{Jain:2004}.\\
The visual interpretation of images is strongly related to the phase of the underlying signal~\cite{Xiao:2004}. Such that the image features ({e.g.}~edges, corners, etc.) occur at parts of the image where the Fourier components are maximally in phase with one another~\cite{Xiao:2004}. Based on local phase information, a research group~\cite{Hacihaliloglu:2009} has presented a robust method for bone surface detection. They use 2-D Log-Gabor filters to derive the phase symmetry (PS) measure a ridge detector for bone localization and automatic segmentation of bone surfaces in ultrasound images. This  technique detects the major axis of  symmetry  of the  signals, and its performance  may  degrade with  the  performance of the reconstruction  method.\\
 In standard medical ultrasound  the images are  reconstructed based  on  the  DAS  beamforming technique. In  this technique received signals from active channels are dynamically delayed and summed in the beamformer. In this case, the achievable resolution, sidelobes level and contrast are limited. Instead, using an adaptive method, such as minimum variance (MV) based beamforming techniques, can enhance the image quality as a result of lower sidelobes, a narrower beamwidth, and superior definition of edges~\cite{Synnevag:2009}. In the MV approach, for each time sample, the delayed received signal from each element is weighted adaptively before summing up in the beamformer. This approach was initially developed by Capon for passive narrow-band applications~\cite{Capon:1969}.\\ 
Several researchers have previously investigated the MV approach in medical ultrasound. They have reported appreciable enhancements in the resolution and contrast in comparison with DAS beamforming~\cite{Wang:2005,Vignon:2008,Synnevag:2009,Asl:2010}. Further, in a simulation study~\cite{Asl:2010} an eigenspace-based MV (ESMV) technique has been employed in order to improve the contrast of the MV beamforming in medical ultrasound imaging. This technique has been developed based on earlier studies in radar imaging~\cite{Cheng:1997,Feldman:1994}. Previous work by our group~\cite{Mehdizadeh:2012a} has demonstrated that in bone imaging scenarios, the robustness of the MV beamformer degrades due to a poor estimation of the covariance matrix. The forward backward (FB) averaging technique  has been  proposed in order to enhance the covariance matrix estimation against signal misalignment due to the shadowing~\cite{Mehdizadeh:2011}. More recently, we have investigated the potential of an ESMV beamforming technique to enhance the edges of the acoustically hard tissues~\cite{Mehdizadeh:unpublished}. We have also shown that by reducing the signal subspace rank the  bone edges   are improved~\cite{Mehdizadeh:unpublished}. Since  the rank estimation is a challenge in ESMV beamformers, in this study we show that the use of a rank one signal subspace can reasonably well preserve the vertebra anatomy and enhance the bone edges in spinal imaging. The constructed images may be  less appealing  from  a visual perspective, but the goal here is to achieve advantages for  post-processing  methods   such as  phase symmetry. In simulation, {in-vitro}, and {in-vivo} studies, we demonstrate that the extracted surfaces from the rank-1  ESMV images  are sharper, and the anatomy of the spine is better defined  in comparison  with their corresponding DAS images.\\
The rest of this paper is organized as follows: in the next section, we first review the beamformer techniques, and the phase symmetry ridge detection method that are employed in this study; then, simulation and experimental setups are introduced. We present the results from simulated data of a point scatterer and vertebra phantoms, followed by results from CT-US registration of a vertebra phantom, and {in-vivo} images of the spine. This section is followed by the discussion on the results.\\


\section{methods}      

\subsection{Minimum variance beamformer}
The minimum variance beamformer employs an element  weight vector which minimizes the variance of the beamformer output under the constraint that the signal arriving from a point of interest is unaffected by the beamformer. In this method, the optimized weights are estimated as:

\begin{equation}
{{\bf{w}}} = \frac{{{{{\bf{R}}}^{ - 1}}{\bf{a}}}}{{{{\bf{a}}^H}{{{\bf{R}}}^{ - 1}}\,{\bf{a}}}},
	\label{eq:eq1}
\end{equation}
where $\textbf{R}$ is the spatial covariance matrix, $\textbf{a}$ is the steering vector, and $(.)^H$ stands for Hermitian transpose. A common estimator for the data covariance matrix is the sample covariance matrix. Therefore, using a method called subarray technique~\cite{Synnevag:2009}, the sample covariance matrix is estimated as: 
\begin{equation}
 {\bf{\hat R}} = \frac{1}{{( {2K + 1})( M -  L + 1)}} \cdot   
  \sum\limits_{ {k =  - K}}^ {K} {\sum\limits_{ {l} = 1}^{M- L+1} {{{{\bf{\bar X}}}_ {l}}[ {n - k}]\,{{{\bf{\bar X}}}_ {l}}{{[ {n - k}]}^{H}}} } , 
 \label{eq:eq2}
\end{equation}

\noindent where

\begin{equation*}
{{\bf{\bar X}}_ {l}}[n] = {\left[ {\begin{array}{*{20}{c}} 
   {{ {x}_ {l}}[n]} & {{ {x}_{ {l} + 1}}[ {n}]} &  \ldots  & {{ {x}_{ {l} + L - 1}}[ {n}]}  \\
\end{array}} \right]^T.}
\end{equation*}

The sample covariance matrix has dimension, $L$, the subarray length, $x_m[n]$ is a time sampled signal from element $m$ of a uniformly spaced linear array with $M$ elements, and $(.)^T$ is transpose operator.  In general, there is a time averaging over index $k$ which has been found to be necessary in order to get proper speckle statistics in the image~\cite{Synnevag:2009}. The subarray technique can be combined with forward-backward averaging to improve the covariance matrix estimation~\cite{Mehdizadeh:unpublished}. The new estimate is expressed as:
\begin{equation}
{{\bf{\hat R}}_{FB} = \frac{1}{2}({\bf{\hat R}} + {\bf{J}}{{\bf{\hat R}}^*}{\bf{J}}),}
\label{eq:eq3}
\end{equation}
\noindent where $\textbf{J}$ is an exchange matrix, the left/right flipped version of the identity matrix, with the same dimension as $\bf\hat R$ , and  $\bf\hat R^*$ denotes the complex conjugate of $\bf\hat R$ . Substituting ${\bf R}$ with either ${\bf\hat R}$  
or ${{\bf\hat R}_{FB}}$  in (\ref{eq:eq1}), the beamformer output is obtained as a coherent average over subarrays by:
\begin{equation}
{\hat {z}}[n] = \frac{1}{M -L + 1}\sum\limits_{ l = 1}^{M -L+1} {{{\bf{w}}^H}}{{{\bf{\bar X}}}_ {l}}[ n],
\label{eq:eq4}
 \end{equation}
 where, $\bf w$ is a vector of time varying complex weights of size $L$. Also, in order to enhance the robustness of the MV estimate a term, $\Delta / {L} \cdot tr\left\{ {\bf{R}} \right\}$ , is added to the diagonal of the covariance matrix before evaluating (\ref{eq:eq1})~\cite{Featherstone:1997}. There are many details about MV beamforming algorithms applied to medical ultrasound imaging, which have been addressed in previous publications~\cite{Wang:2005,Vignon:2008,Synnevag:2009,Asl:2010}. In this paper we use the method that is described in~\cite{Synnevag:2009}.
 
\subsection{Eigenspace-Based beamformer}      
The eigenspace-based beamformer (ESMV) utilizes the eigen structure of the covariance matrix to estimate MV weights~\cite{Feldman:1994, Chang:1992}. With assumption of $j\leq L$ , the sample covariance matrix ${\bf\hat R}$  defined by (\ref{eq:eq2}) is eigendecomposed as:   

\begin{equation}
{\bf{\hat R}} = {\bf{E\Lambda E}}_{}^H = {{\bf{E}}_s}{{\bf{\Lambda }}_s}{\bf{E}}_s^H + \,{{\bf{E}}_N}{{\bf{\Lambda }}_N}{\bf{E}}_N^H,
\label{eq:eq5}
 \end{equation}
\noindent where

\begin{equation}
\begin{split}
 &{\bf{E}_{\rm{s}}} =[{{\bf{e}}_1},...,{{\bf{e}}_{\it j}}],\,\,\,\,\,\,\,\,\,\,\,\,\,\,{\bf{E}_{\rm{N}}} = [{{\bf{e}}_{{\it j} + 1}},...,{{\bf{e}}_{{\it L}}}],\\ 
 &{\bf{\Lambda}_{\rm{s}}} = {\it diag}[{\it{\lambda _{\rm 1}},...,{\lambda _j}}],\,\,{{\bf{\Lambda}_{\rm{N}}}} = {\it diag}[{\it{\lambda _{j + {\rm 1}}},...,{\lambda _{L }}}],
 \label{eq:eq6}
 \end{split}
\end{equation}

\noindent and, ${{\lambda _1} \ge {\lambda _2} \ge .... \ge {\lambda _{L }}}$ are eigenvalues in descending order, and ${\bf {e}}_l$, $\it l ={\rm {1}},..., L$ are the corresponding orthonormal eigenvectors. We refer to the subspace spanned by the columns  of ${\bf{E}}_s$ as the signal subspace and to that of ${\bf{E}}_N$ as the noise subspace. Ideally, the direction of the steering vector and the noise subspace are orthogonal, {i.e.}~${\bf{E}}_N^H{\bf{a}} = 0$~\cite{Chang:1992}. This  will result in a weight vector as~\cite{Mehdizadeh:unpublished,Feldman:1994}:

 \begin{equation}
{{{\bf{w}}_p} = {{\bf{E}}_s}{\bf{E}}_s^H{\bf{w}}.}
\label{eq:eq7}
\end{equation}
Equation~\ref{eq:eq7} can be interpreted as the projection of $\bf w$ on the signal subspace of ${\bf\hat R}$~\cite{Feldman:1994}. We select the rank of the signal subspace employing the cross-spectral metric~\cite{VanTrees:2002}. 
The output signal power of the minimum variance beamformer can be expressed based on the cross-spectral metric as given in chapter 6.8.2 of~\cite{VanTrees:2002}.

\begin{equation}
{{\sigma^{2}_{z}} = ({{\bf{a}}^H}{\bf{E}}{\bf{\Lambda}^{-1} }{{\bf{E}}^H}{\bf{a}})^{-1} =( \sum\limits_{k = 1}^L {\frac{{\rho _k^2}}{{{\lambda _k}}}})^{-1} ,\,\,\rho  = {{\bf{E}}^H}{\bf{a}},}
\label{eq:eq8}
\end{equation}

\noindent where ${{{\rho _k^2} \mathord{\left/ {\vphantom {{\rho _k^2} {{\lambda _k}}}} \right. \kern-\nulldelimiterspace} {{\lambda _k}}}}$ is the cross-spectral metric for the $k^{th}$ eigenvalue. We select the rank of ${\bf{E}}_s$ by identifying the $j$ largest eigenvalues for which the sum of their cross-spectral metric is $\beta$ times smaller than the total output signal power~(${\sigma^{2}_{z}}$)~\cite{Mehdizadeh:unpublished}.

\subsection{Ridge detection}
The images were obtained from different beamforming techniques using Matlab (the Mathworks, Natick, MA, U.S), and resampled to $512\times512$ isotropic pixels to form the basis for further image processing by phase symmetry filtering. We also implemented the phase symmetry algorithm in Matlab. A log-Gabor filter was defined in polar coordinates as the product of a radial factor by an angular factor:

\begin{equation}
f(r,\theta ) = \exp \left[ { - \frac{1}{2} \cdot {{\left( {\frac{{\log (\frac{r}{{{r_0}}})}}{{\log (\frac{{{\sigma _r}}}{{{r_0}}})}}} \right)}^2}} \right] \cdot \exp \left[ { - \frac{1}{2} \cdot {{\left( {\frac{{\alpha (\theta ,{\theta _0})}}{{{\sigma _\alpha }}}} \right)}^2}} \right],
\label{eq:eq9}
\end{equation}
where $r$ and $\theta$ are the coordinates in the Fourier-transformed image, $r_0$ the characteristic radius, and $\sigma_r$ the radial standard deviation.  For the radial part, we choose  empirically  $\sigma_r/r_0$ as  0.15 . The angular factor $\alpha(\theta,\theta_0)$, shows the angle between the position vector and the direction of the filter, and $\sigma_\alpha$ is angular standard deviation that is assumed to be $\pi/6$  in this study. The two-dimensional Fast Fourier Transform $(F)$ of the image is multiplied by the filter and their product is inversely transformed by $F^{-1}$.\\
A bank of filters is used with different $r_0$ and $\theta_0$ in order to enhance features of the image of different sizes and orientations. For each image, we use filters consisting of the combinations of several characteristic radius $r_0$ exponentially distributed from $2^{-5}$ to $2^{-9}$ in pixel space, and 6 characteristic orientations $\theta_0$~(0, $[-\pi/2+k\cdot\pi/6$ , $k = 0,\pm1,\pm2$]), distributed around the main direction of the ultrasound beam (downward). The filters and the corresponding filtered images were marked by the index $i$.\\
At each point of the imaging field, the real and imaginary parts of the filtered images are combined to form a metric of the phase symmetry (PS):
\begin{equation}
PS = \frac{{\sum\nolimits_i {\left\lfloor {({e_i} - \left| {{o_i}} \right|) - T_n} \right\rfloor } }}{{\sum\nolimits_i {\sqrt {e_i^2 + o_i^2}  + \varepsilon } }},
\label{eq:eq10}
\end{equation}
\noindent where $\left\lfloor u \right\rfloor$  denotes $max(u,0)$ and $e_i$ and $o_i$ are the even and odd (real and imaginary) part of the image processed by filter $i$, $T_n$ is a noise threshold (dimensionless) and $\epsilon$ is included simply to avoid division by zero ($\epsilon ~=~10^{-10}$). The asymmetrical treatment of even and odd components reflects a polarity choice where only dark-to-light-to-dark features are detected. There are more details about the PS method which have been addressed in previous publications~\cite{Hacihaliloglu:2009,Hacihaliloglu:2011a}.\\
The    threshold, angular and radial standard deviations are chosen empirically to provide images with the least noise; yet retaining the most information.
 They are maintained identical for all images. The central radial frequency combinations are adjusted to best fit different applications. For the patient imaging of the sagittal lamina  view, we used $2^{-8}$ and $2^{-9}$, for simulations, the   sagittal spinous process  view, and  the transversal lamina view, $2^{-7}$ to $2^{-9}$, and for the water bath images we used, $2^{-5}$ to $2^{-9}$. Further,  we set the noise threshold $T_n =10$ for    the water bath images, and $T_n =15$ for the rest of  images.

\subsection{Simulation setup}
\begin{figure*}[!htb]
\centering
\includegraphics[width=5.0in]{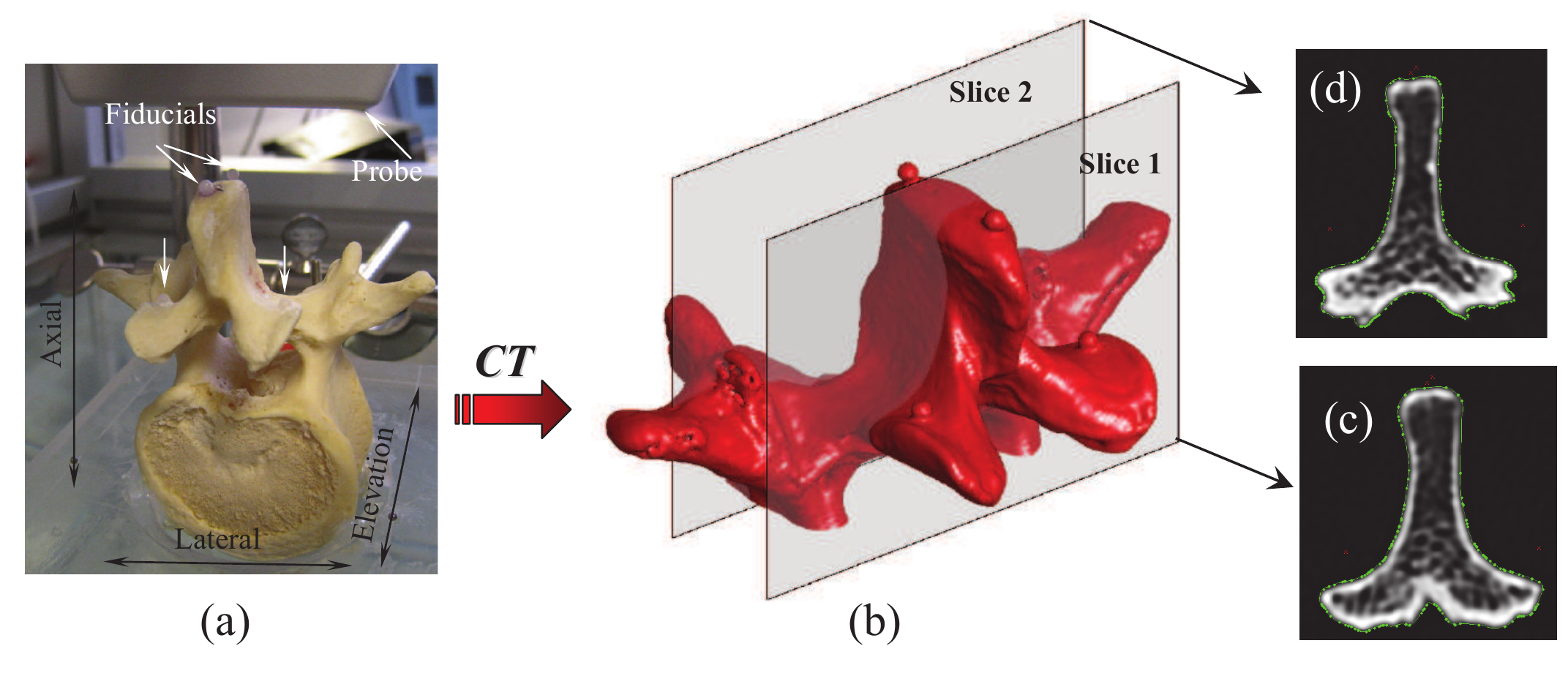}

\caption{(a) An illustration of the vertebra phantom, (b) 3D model constructed from the CT dataset  and projection of the selected slices, (c) CT image of slice 1 and its corresponding surface profile, and (d) CT image of slice 2 and its corresponding surface profile.} 
\label{fig1}
\end{figure*}

 In this study, we simulate two different phantoms using Field II~\cite{Jensen:1996}: a single point scatterer phantom, and a vertebra phantom.
The vertebra phantom consists of a vertebra body that is embedded in the soft tissue. We use the simulation scenario proposed in~\cite{Mehdizadeh:2012a} for the vertebra phantom. We assume that the bone structure is completely attenuating. Therefore, it shadows point scatterers and surfaces which are not directly visible to the imaging aperture. The 3D geometry of the vertebra body is obtained by CT scanning of a human lumbar vertebra specimen (Fig.~\ref{fig1}). By utilizing Matlab and VTK (Kitware, New York, NY, U.S) the 3D vertebra dataset has been segmented into triangular  surfaces~\cite{Lorensen:1987}. Then,  equally weighted  and spaced point scatterers are generated on the triangulated surfaces with a concentration of 200 scatterers/mm$^2$. The soft tissue is modeled by $1.0\times10^6$ equal amplitude point scatterers that are uniformly distributed in a region of $30\times6\times�25$ mm$^3$. The number of scatterers per resolution cell exceeds 10, which is recommended to simulate speckle~\cite{Wagner:1988}. The scatterers that are inside the vertebra body are identified and removed from the phantom. The image of the shadowed surfaces and point scatterers are modified by introducing a binary apodization-based shadowing model~\cite{Mehdizadeh:2012a}. This model is applied to Field II in order to make an image of the vertebra phantom.\\
We simulate images employing a linear array with 128 elements and a center frequency of 5 MHz ($f_0$) with 60 percent $-6$ dB fractional bandwidth. The array's elevation focus is 19 mm, and its pitch equals 0.308 mm. The maximum accessible aperture size for this array transducer is 38.70 mm ($M$~=~128). The array is excited by 1.5 periods of a square wave at the center frequency of the array. In all simulations, a beam density of 1 beam per element, a fixed transmit focus, and dynamic receive focusing is used. In addition, the f number in the transmit is set to  FN$_{TX}$~=~2.8, while the receive f number is set to FN$_{RX}$~=~2.5 for the point scatterer phantom, and FN$_{RX}$~=~1.5 for the vertebra phantom. We select a large FN$_{RX}$ for the point scatterer phantom imaging scenario in order to achieve a wide enough beam width to ease further analysis. The transmit focal depth is set to 15 mm unless otherwise specified. The channel data are acquired for each scan line with a sampling frequency of 100 MHz. For all beamformers after applying delays the channel data are down-sampled to 20 MHz. We computed the analytic signals by applying the Hilbert transform to the channel data. Consequently, in the DAS approach the delayed received channel data are summed up for each scan line, without any apodization, whereas for MV-based beamformers the optimal aperture weights are estimated for each time sample before summation. In the adaptive approaches, we use diagonal loading with  $\Delta~=~5\%$ in all simulations. 

\subsection{Experimental setup}

We have 2 different experimental cases: registration of a single vertebra, and imaging the spine  of a volunteer. In the first experiment we use a human lumbar vertebra specimen (L3) and align the 3D-CT dataset to the 3D-US one. Thus, we secure the vertebra specimen in a rigid holder and glue 4 small plastic balls (fiducials) with a diameter of 2 mm on the vertebra body; two on the spinous process (top) and two on the lamina. Positions of these fiducials are illustrated in Fig.~\ref{fig1}(a). \\
The 3D US volume is constructed from 2D US slices acquired from imaging the vertebra specimen in a water bath, and by moving the probe using a 2D robot in the elevation direction  by  a step of 0.5 mm [Fig.~\ref{fig1}(a)]. The constructed 3D-US volume consists of $512\times512\times61$ voxels with a dimension of 0.077 mm$\times$0.077 mm$\times$0.5 mm. Subsequently a CT dataset of the vertebra specimen is prepared using a high-resolution CT imager (Siemens, Somantom Definition Flash). This results in a CT volume of $512\times512\times374$ voxels with a resolution of (0.19 mm$\times$0.19 mm$\times$0.3 mm). 
For registration, the coordinates of the fiducials' tip are manually selected both in US and CT datasets. A landmark-based rigid registration algorithm is used to transform the CT dataset in order to match the 3D-US volume. The CT slices are resampled to the in-plane US resolution. Since CT-US registration is performed, the bone iso-surfaces are extracted from the CT volume employing the Marching cubes algorithm in VTK~\cite{Lorensen:1987}. This is expected to match the ones in US and can be used as the gold standard (GS) reference. An  empirically  chosen thresholding value of -524 Hounsfield unit (HU) is used to extract the surface profile from the CT slices.  We   measure the  registration accuracy by calculating  the fiducial registration error (FRE) \cite{Fitzpatrick:2009}.\\ 
To compare our images with the gold standard CT  surface profile, a signed distance distribution of the US intensity  values  is computed~\cite{Hacihaliloglu:2011b}. First, US images   are mapped to   their  corresponding  CT image, and  the  normal distance of non-zero intensity pixels are computed  with  respect to the  extracted GS profile. The  pixels located inside the GS profile  have  positive  and  the  pixels located  outside  the GS profile have negative distance values. This  produces  a   set  of  intensity/ signed distance  pairs. The  high  intensity values around the zero  distance indicates  the  bone localization  accuracy, and   the  concentration  of the intensity values in  positive/negative  distances  shows  the noise level inside/outside of the bone  surface.\\        
In the {in-vivo} experiments, we use a male healthy volunteer. His lumbar vertebra (L2) is scanned in three different planes: sagittal plane of the spinous process, and sagittal  and transversal planes  of the lamina. For scanning the spinous process, we use a 10 mm stand-off (SonarAid, Wolhusen, Lucerne, Switzerland) in order to improve the matching between probe and skin. The scans were preformed after obtaining signed consent from the volunteer.\\
In the experimental studies, channel data are acquired using a SonixMDP scanner (Ultrasonix medical corporation, Vancouver, British Columbia, Canada), along with a linear array transducer (L14-5/38) with 128 elements, centre frequency of 5 MHz, and pitch of 0.308 mm. We use 256 imaging beams which are transmitted with FN$_{TX}$~=~2.8, and received with FN$_{RX}$~=~1.5. Further, the receive aperture walks with the transmit aperture, meaning that the active receive elements are centered on the transmit beam axes. The SonixDAQ (Ultrasonix medical corporation, Vancouver, British Columbia, Canada) is used to capture the channel data. This module allows us to store RF data acquired from 128 elements simultaneously. For the beamforming, the channel data related to each beam is first determined and delayed. Then, the ESMV beamforming method is applied to construct images of interest. As for the simulations, $\Delta~=~5\%$ is used for the diagonal loading purpose. Further, after construction of the images a 2D median filter with a window size of $3\times3$ is applied to smooth the images. 

\section{Results}
\label{Results}
\subsection{Effects of the largest eigenvalue on image of a point scatterer:}
\begin{figure*}[!htb]
\centering
\includegraphics[width=3.0in]{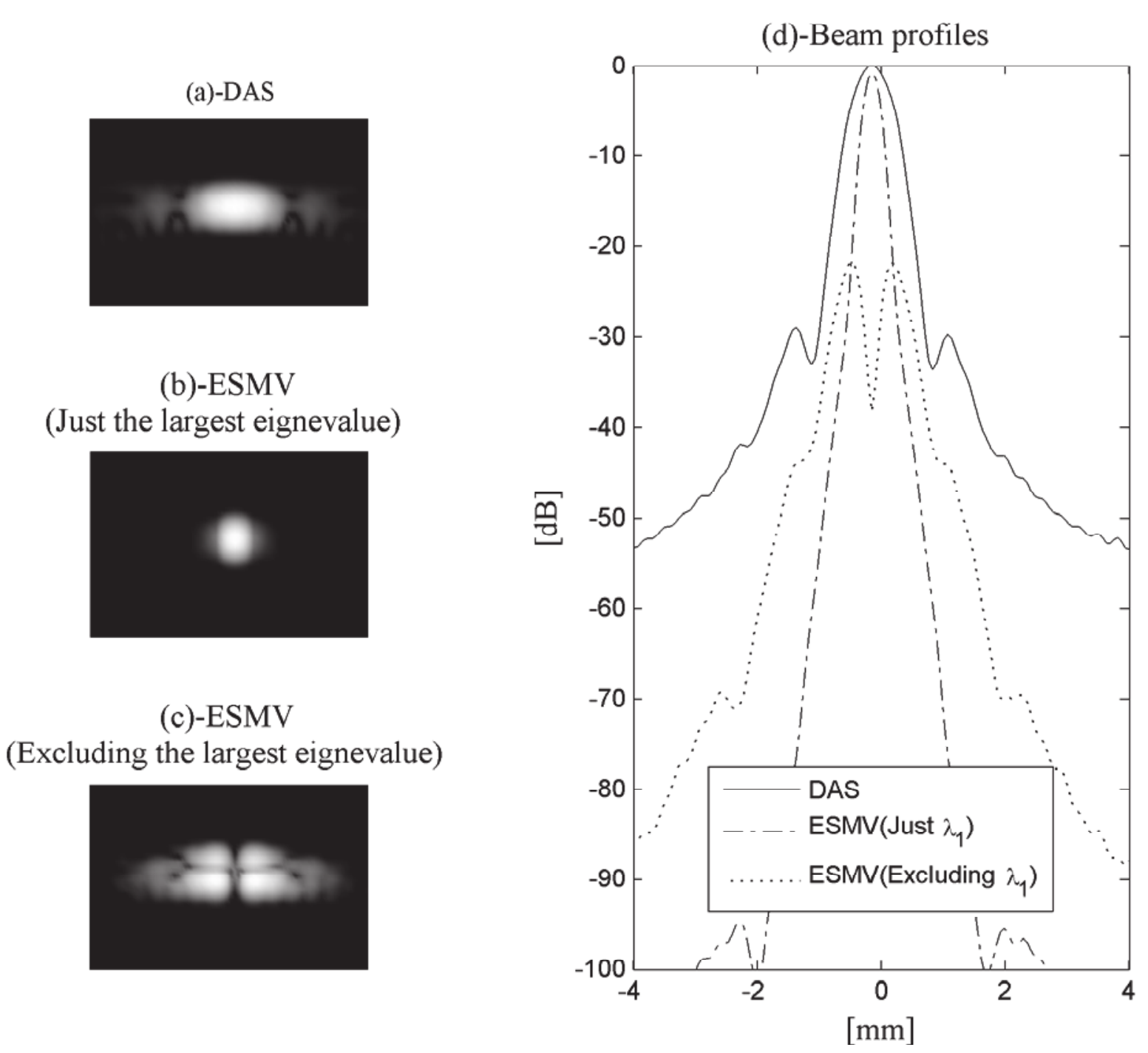}
\caption{Simulated point scatterer using a 128 element, 5 MHz transducer. The point scatterer is located at the transmit focal depth which is 15 mm. Dynamic focusing is used for the received beams. (a) DAS, (b) ESMV (just the largest eigenvalue), (c) ESMV (excluding the largest eigenvalue), (d) two-way beam profiles corresponding to the images in (a)-(c). The dynamic range is 50 dB. $L=M/2$, $K=0$ are used for (b)-(c). In (a)-(c) image dimensions are 3 mm$\times$5 mm (depth$\times$lateral).}
\label{fig2}
\end{figure*}

Fig.~\ref{fig2} demonstrates the effect of using only the largest eigenvalue on the image of a point scatterer. Fig.~\ref{fig2}(a) shows the DAS image of the simulated point scatterer. Fig.~\ref{fig2}(b) presents the ESMV image when only the largest eigenvalue is used for estimating the signal subspace ($\bf{E}_s$). In comparison with Fig.~\ref{fig2}(a), the point scatterer is defined with higher resolution and the sidelobe level is decreased. In Fig.~\ref{fig2}(c), it is assumed that all eigenvalues contribute in the signal subspace except the largest one ($\lambda_1$). In this scenario, the image of the point scatterer is completely distorted. In Fig.~\ref{fig2}(d) the beam profiles corresponding to Figs.~\ref{fig2}(a)~-~(c) are compared. In this figure it can be seen that using $\lambda_1$ in the ESMV beamformer results in a -12 dB beamwidth of 0.35 mm. This value is about 0.8 mm for DAS. Ideally, the sidelobe levels are decreased from -30 dB in DAS to -95 dB for ESMV. Also, it can be seen that when $\lambda_1$ is excluded a major part of the mainlobe between 0 and -20 dB is removed [Fig.~\ref{fig2}(d)]. 

\subsection{Simulated vertebra phantom:}
\begin{figure*}[!htb]
\centering
\includegraphics[width=6in]{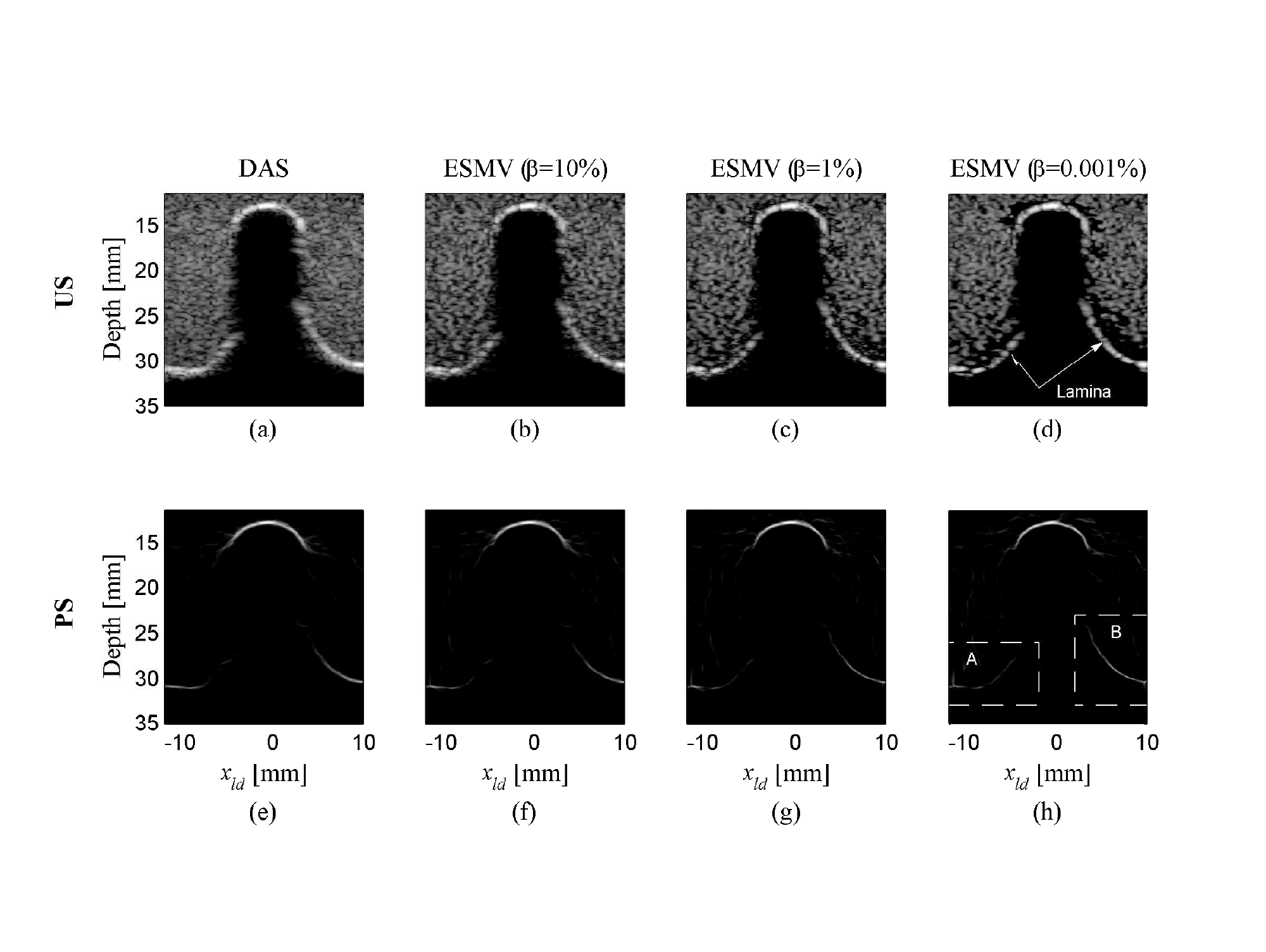}
\caption{Simulated vertebra phantom using a 128 element, 5 MHz transducer. The transmit focal depth is 15 mm and dynamic focusing is used for the received beams. (a) DAS (US), (b) ESMV (US, $\beta=10\%$), (c) ESMV (US, $\beta=1\%$), (d) ESMV (US, $\beta=0.001\%$) (e) DAS (PS), (f) ESMV (PS, $\beta=10\%$), (g) ESMV (PS, $\beta=1\%$), and (h) ESMV (PS, $\beta=0.001\%$). The dynamic range is 60 dB for (a)-(d). $L=M/2$, $K=10$ are used for (b)-(d).}
\label{fig3}
\end{figure*}

Fig.~\ref{fig3} shows simulated images of the vertebra phantom introduced in the simulation setup section and its corresponding phase symmetry images for different beamformers. In this imaging scenario, the transmit focal depth is  15 mm. Fig.~\ref{fig3}(a) shows the DAS image of the vertebra phantom. Figs.~\ref{fig3}(b)~-~(d) present ESMV images for different eigenvalue threshold values ($\beta$). In Fig.~\ref{fig3}(d), $\beta=0.001\%$ is selected to ensure that just the largest eigenvalue is used. It can be seen that by decreasing $\beta$ the speckle pattern in the neighboring region of the vertebra body is distorted, and for $\beta=0.001\%$ it is almost removed, especially between the depths of 25 mm and 32 mm. This effect can be partly seen around the spinous process~(top of the vertebra) at a depth of 15 mm. Figs.~\ref{fig3}(e)~-~(h) show PS images related to Figs~\ref{fig3}(a)~-~(d). It can be seen that by decreasing $\beta$  the  bone  boundaries become  sharper. In Fig.~\ref{fig3}(h),  we observe  that  a larger segment  of the lamina is detected   between a depth  of 26 mm  and  29 mm in box A, and   between a depth of 23 mm  and 26 mm in  box B  in comparison  with Fig.~\ref{fig3}(e). 

\subsection{Registration of a vertebra:}
\begin{figure*}[!htb]
\centering
\includegraphics[width=4.5in]{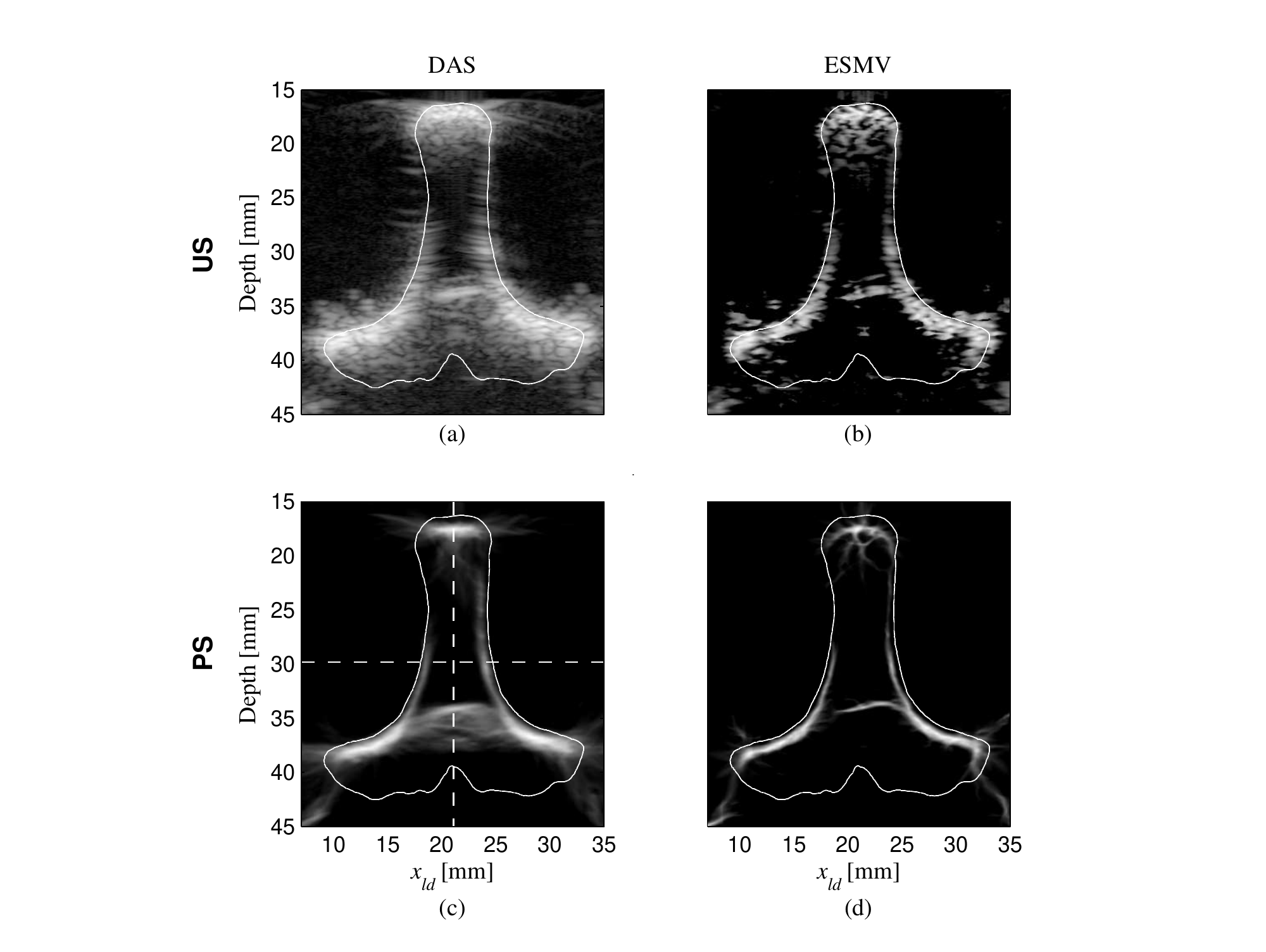}
\caption{Registration of the CT image to ultrasound image slice 1 for DAS and ESMV images and their corresponding PS images, (a) DAS (US), (b) ESMV (US, $\beta=0.001\%$ ), (c) DAS (PS), and (d) ESMV (PS, $\beta=0.001\%$). The dynamic range is 70 dB for (a) and (b). $L=M/2, K=0$ are assumed for (b).}
\label{fig4}
\end{figure*}

\begin{figure*}[!htb]
\centering
\includegraphics[width=4.5in]{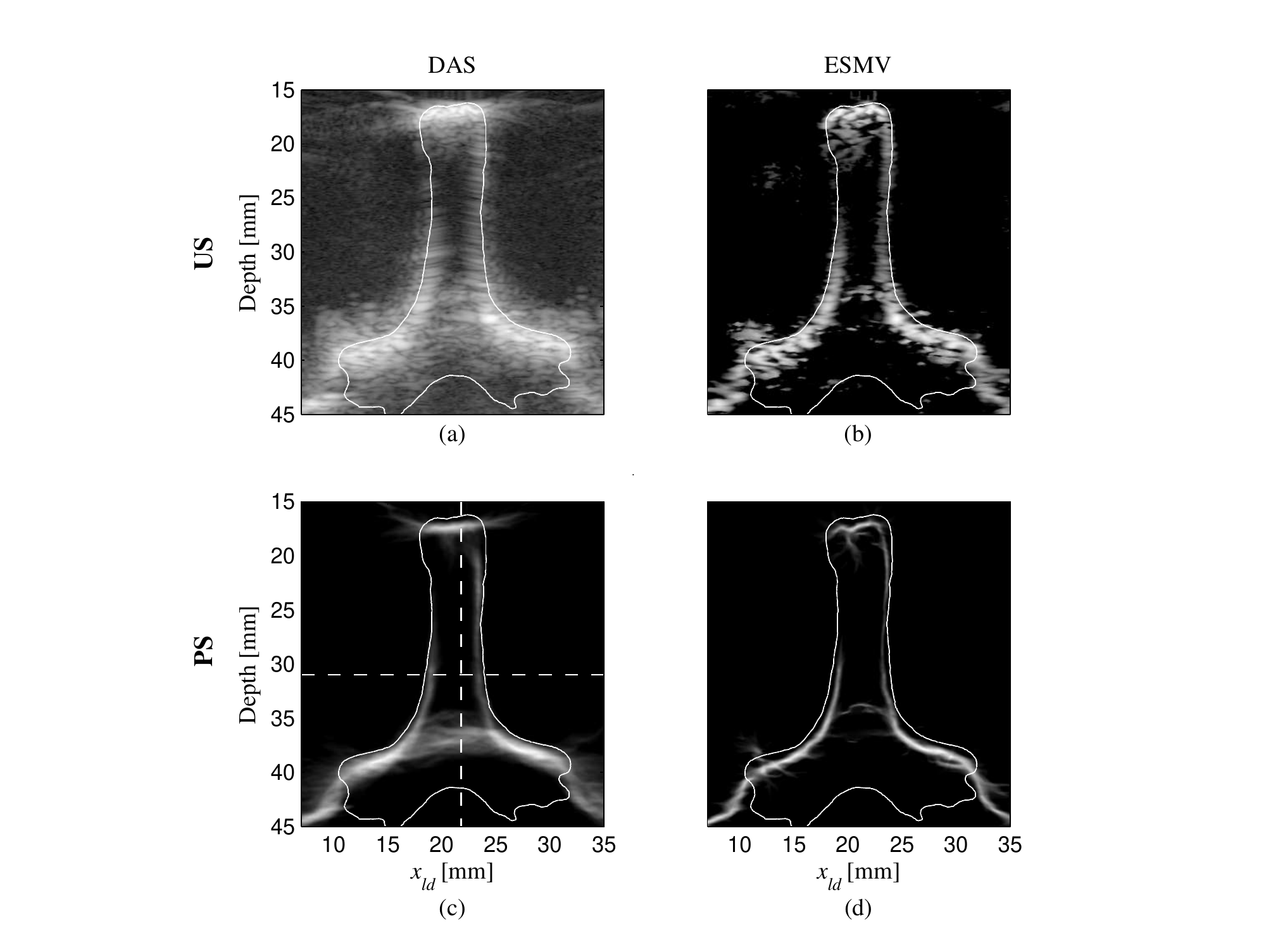}
\caption{Registration of the CT image to ultrasound image slice 2 for DAS and ESMV images and their corresponding PS images, (a) DAS (US), (b) ESMV (US, $\beta=0.001\%$ ), (c) DAS (PS), and (d) ESMV (PS, $\beta=0.001\%$). The dynamic range is 70 dB for (a) and (b). $L=M/2, K=0$ are assumed for (b).}
\label{fig5}
\end{figure*}

 Figs.~\ref{fig4} and ~\ref{fig5} show the CT gold standard surface profile overlaid on the ultrasound images for the two different vertebra slices of Fig.~\ref{fig1}(b). In this registration setup, the FRE   value  is  calculated as  0.13 mm. Figs.~\ref{fig4}(a)~and~(b) show the DAS and ESMV images of slice~1. The CT profile matches well on outer boundary of the vertebra in both images. In the DAS image [Fig.~\ref{fig4}(a)] the sidelobe noise is clearly observed around the spinous process between a depth of 15 mm and 20 mm. Also, the sidewall boundaries are stretched due to the shadowing effect~\cite{Mehdizadeh:2012a}, whereas in the ESMV image the sidelobe noise is decreased and the boundaries are enhanced. In Figs.~\ref{fig4}(c)~and~(d) a deviation of the surface from the gold standard surface is observed, particularly on spinous process (top of the vertebra). In Fig.~\ref{fig4}(c) the curvature of spinous process profile has been distorted, whereas the anatomy of the vertebra is  preserved reasonably well in Fig.~\ref{fig4}(d). Further,  the acoustical noise  observed  inside the bone, between  a depth of 35 mm  and 40 mm, is  reduced in  Fig.~\ref{fig4}(d) than that of  Fig.~\ref{fig4}(c).\\

Figs.~\ref{fig5}(a)~and~(b) show the DAS and ESMV images of slice 2, and Figs.~\ref{fig5}(c)~and~(d) demonstrate their corresponding PS images. Comparing the  DAS  and ESMV based US images, the  bone  edges are  improved in  Fig.~\ref{fig5}(b) in comparison  with  Fig.~\ref{fig5}(a). Further, comparing to the gold standard surface profile, in Fig.~\ref{fig5}(d) the anatomy of the spinous process is preserved whereas it is distorted in Fig.~\ref{fig5}(c). Also, in Fig.~\ref{fig5}(d), the detected surface is   sharper  in comparison with Fig.~\ref{fig5}(c).\\

\begin{figure*}[!htb]
\centering
\includegraphics[width=5in]{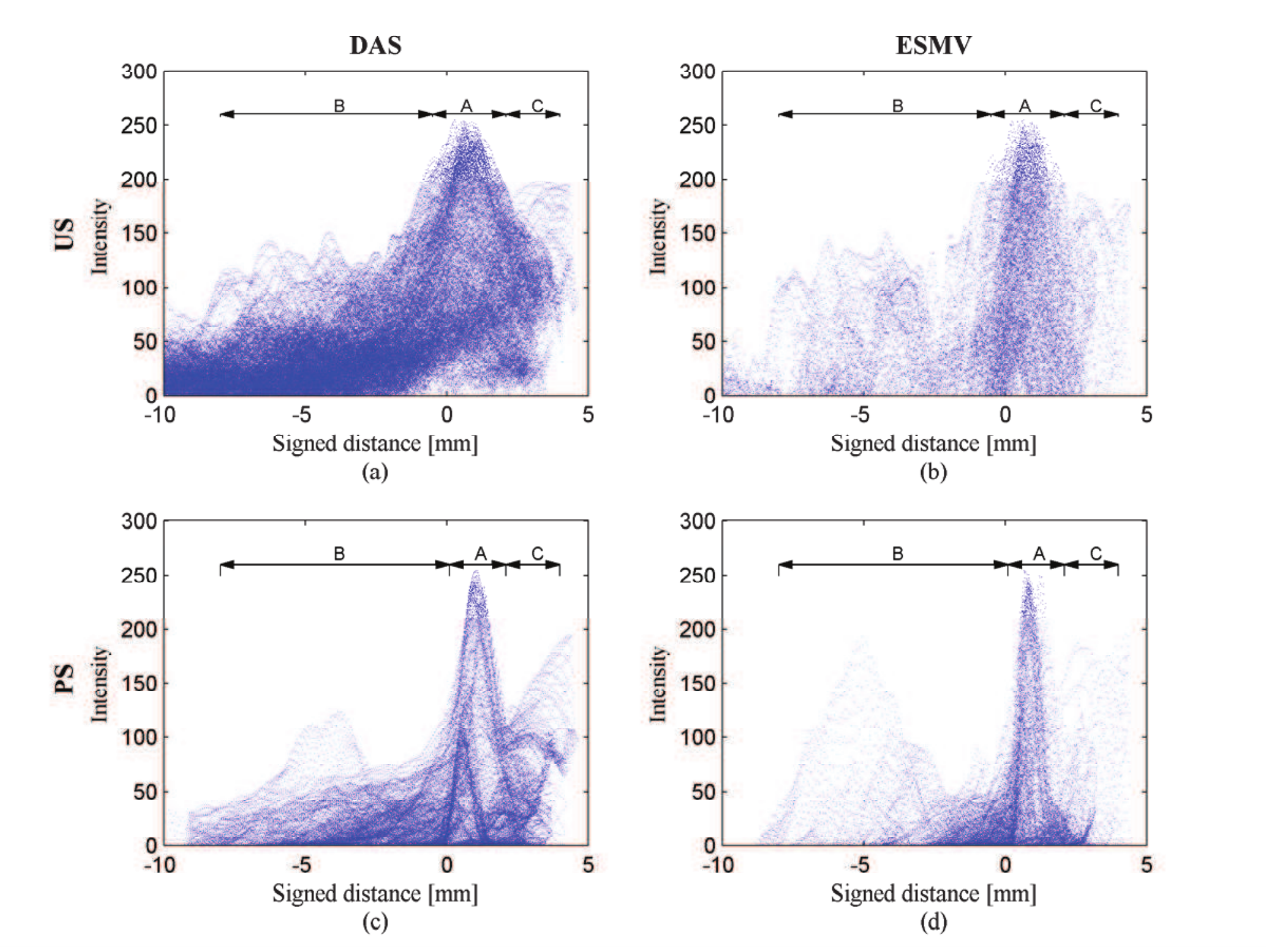}
\caption{Signed distance plots obtained from  Fig.~\ref{fig4}. (a) DAS (US), (b) ESMV (US, $\beta=0.001\%$ ), (c) DAS (PS), and (d) ESMV (PS, $\beta=0.001\%$).}
\label{fig6}
\end{figure*}

Fig.~\ref{fig6} presents the distribution of intensity values and their corresponding signed distances for the images in Fig.~\ref{fig4}. Each graph is  divided  into  three  regions: A , B, and C. The  region A  indicates the intensity distribution around the bone  surface,  defined  between -0.5 mm  and 2.1 mm in US images, and  between 0.1 mm and 2.1 mm in  PS images. Both  US  and  PS images  corresponding  to the ESMV  beamforming technique   have  less  noise level in  the  regions  B  and C (Table.~\ref{tab:TABEL0}),  and a narrower distribution in the  region A in comparison with those of the DAS beamforming technique. Comparing  the  PS  images, the mean surface localization error, calculated in  region A, is  0.90 mm (STD =0.85 mm ) for  DAS  and 0.79 mm (STD = 0.77 mm) for  ESMV.\\

\begin{figure*}[!htb]
\centering
\includegraphics[width=5in]{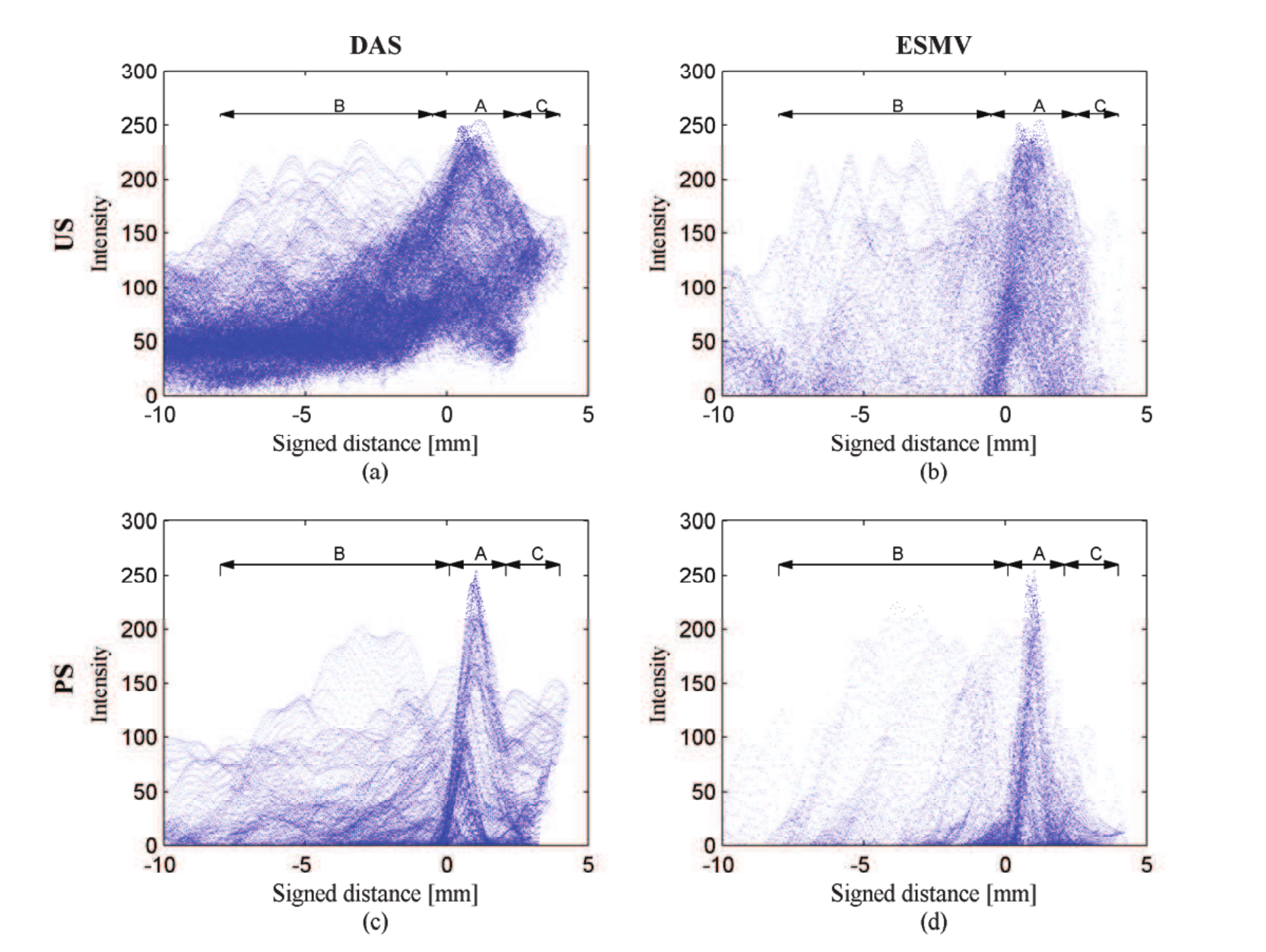}
\caption{Signed distance plots obtained from  Fig.~\ref{fig5}. (a) DAS (US), (b) ESMV (US, $\beta=0.001\%$ ), (c) DAS (PS), and (d) ESMV (PS, $\beta=0.001\%$).}
\label{fig7}
\end{figure*}

 Fig.~\ref{fig7} presents the distribution of intensity values with their corresponding signed distance for the images in Fig.~\ref{fig5}. The regions A, B,  and C are  defined the same as  in Fig.~\ref{fig6}. The  noise levels  in the  regions B and C of the ESMV image are almost 16$\%$  and  13$\%$  of that  in  the DAS image. Comparing  Figs.~\ref{fig6}(c) and (d),  we observe that the concentration of intensity values are  much  less in  regions B, and C in the ESMV image~(PS) than in  the  DAS image~(PS). The mean localization error is 0.97 mm (STD =0.57 mm) for DAS  and  0.95 mm (STD=0.45 mm) for ESMV.

Table \ref{tab:TABEL0}  shows quantitative  results for  the  image  quality  assessment of the  vertebra slices  in Figs.~\ref{fig4}~and~\ref{fig5}. This table shows the  bone surface localization errors and the  noise level  in  the US  and PS images obtained from the DAS and  ESMV beamforming techniques. 
 
\begin{figure*}[!htb]
\centering
\includegraphics[width=5in]{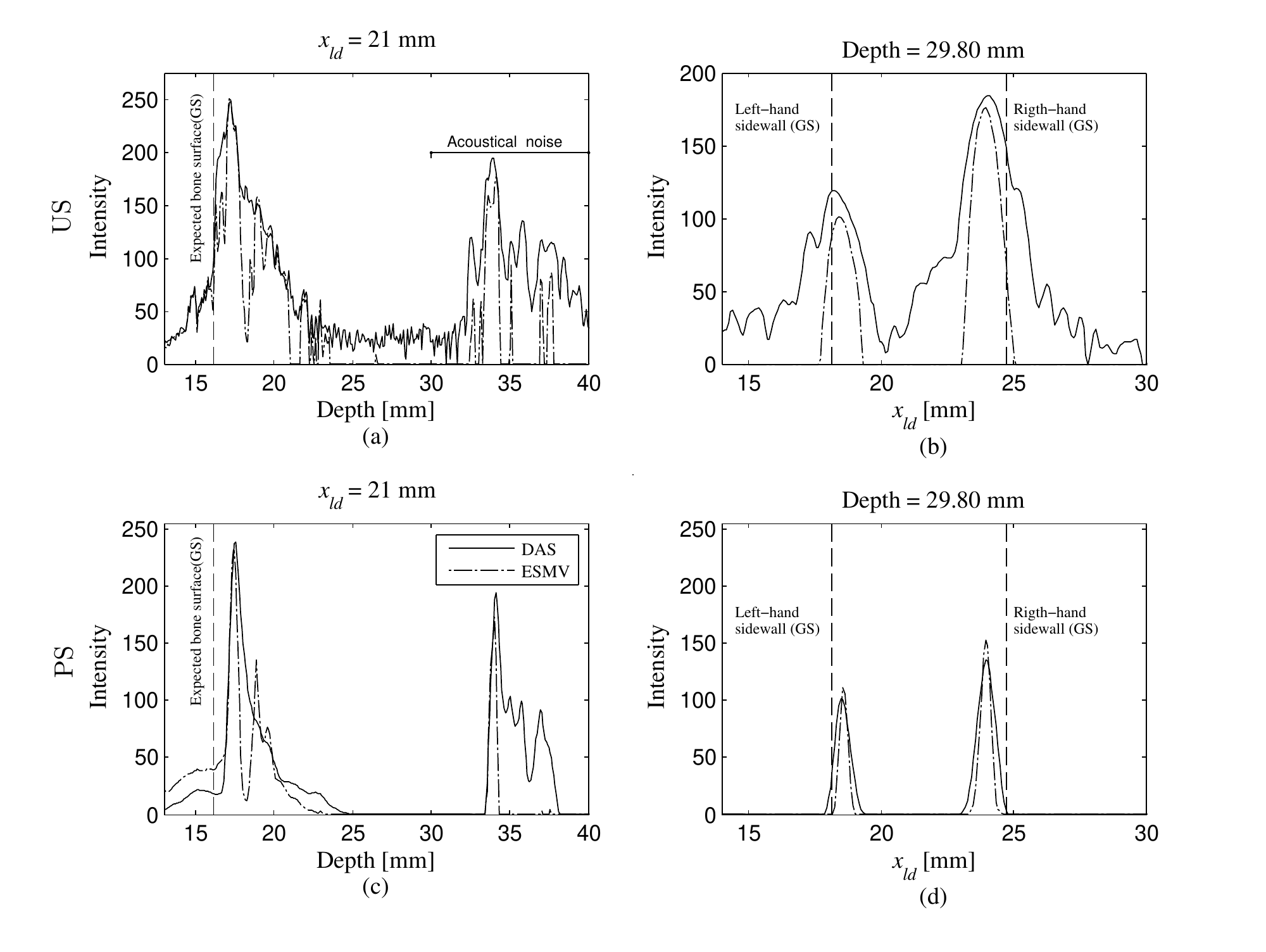}
\caption{Comparison of image lines in Figure~\ref{fig4} for the DAS and ESMV images, (a) US images-$x_{ld}$~=~13.70 mm, (b)US images-$x_{ld}$~=~21 mm, (c)PS-images depth~=~32.10 mm, and (d)PS-images depth~=~36.60 mm.}
\label{fig8}
\end{figure*}

In Fig.~\ref{fig8}, we present two different image lines of the US  and PS  images presented in Fig.~\ref{fig4}. These lines are marked in Fig.~\ref{fig4}(c). In Figs.~\ref{fig8}(a)~-~(d) the location of the bone surface obtained from the gold standard reference is marked by vertical dash-dot lines. In Fig.~\ref{fig8}(a),  there  is a  peak bias of 1.12 mm for both DAS and ESMV, which is measured  relative to  the gold standard reference. The mean intensity of  acoustical noise,  measured between a depth of 30 mm  and 40 mm, is decreased from  69.01 in DAS to  14.33 in ESMV . The  profile widths at an intensity value of  200 are 1.03 mm for DAS and 0.73 mm for ESMV.
Fig.~\ref{fig8}(b) shows a horizontal image line  at  $\rm Depth$~=~29.80 mm. The peaks at  $x_{ld}$~$\approx$~18.30 mm  and  $x_{ld}$~$\approx$~24 mm  indicate the  left-hand and  right-hand sidewalls at the corresponding depth. The profile width  at a pixel intensity of 150 is  0.75 mm for ESMV and 1.39 mm for DAS  for the right-hand sidewall. Fig.~\ref{fig8}(c) presents   the PS scan-lines corresponding to Fig.~\ref{fig8}(a). In  this figure, the profile width measured at an intensity value of 200 is 0.44 mm  for DAS, and 0.30 mm for ESMV. Further, the intensity  level drops  by  129 for DAS  and  by 218 for ESMV between a depth of 17.45 mm  and 18 mm. Also, the mean noise  level  is 24.50  for DAS  and 4.67  for ESMV  between a depth of 30 mm and 40 mm. In  Fig.~\ref{fig8}(d), the profile width  at an intensity level of 100  is  0.51 mm  for DAS and 0.34 mm  for ESMV, measured around the right-hand sidewall. 
 
 \begin{table*}[!htb]
\caption{ Quantitative  results for  the  image  quality  assessment of the  vertebra slices  in Figs.~\ref{fig4}~and~\ref{fig5}. The mean localization error is calculated in region~A. Regions A, B, and  C are presented in Figs.~\ref{fig6}~and~\ref{fig7}}.
\centering
\begin{tabular}{c c c c c c c c c c }
\hline
& \multicolumn{1}{c}{Slice} & \multicolumn{2}{c}{Mean localization error [mm]} & \multicolumn{2}{c}{STD [mm]} 
 & \multicolumn{2}{c}{Mean intensity B} & \multicolumn{2}{c}{Mean intensity C} \\
 &   &   DAS   & ESMV   &  DAS  & ESMV  & DAS & ESMV & DAS & ESMV\\
\hline\
\multirow{2}{*}{US}& 1  &  0.70   & 0.66   & 0.72 & 0.65    & 27.18 & 4.71  & 92.41  & 16.81 \\
& 2  &  0.90   & 0.79   & 0.85 & 0.77    & 63.06 & 10.17 & 125.52 & 15.80 \\
\hline\
 \multirow{2}{*}{PS}& 1  &  0.94 & 0.84 & 0.53 & 0.42 & 32.20& 6.60& 3.06 & 1.24 \\
 & 2  &  0.97 & 0.95 & 0.57 & 0.45 & 32.24& 3.93& 5.16 & 1.88 \\
\hline\
 & \multicolumn{9}{l}{STD = standard deviation} 
\end{tabular}
\label{tab:TABEL0}
\end{table*}

\begin{figure*}[!htb]
\centering
\includegraphics[width=5in]{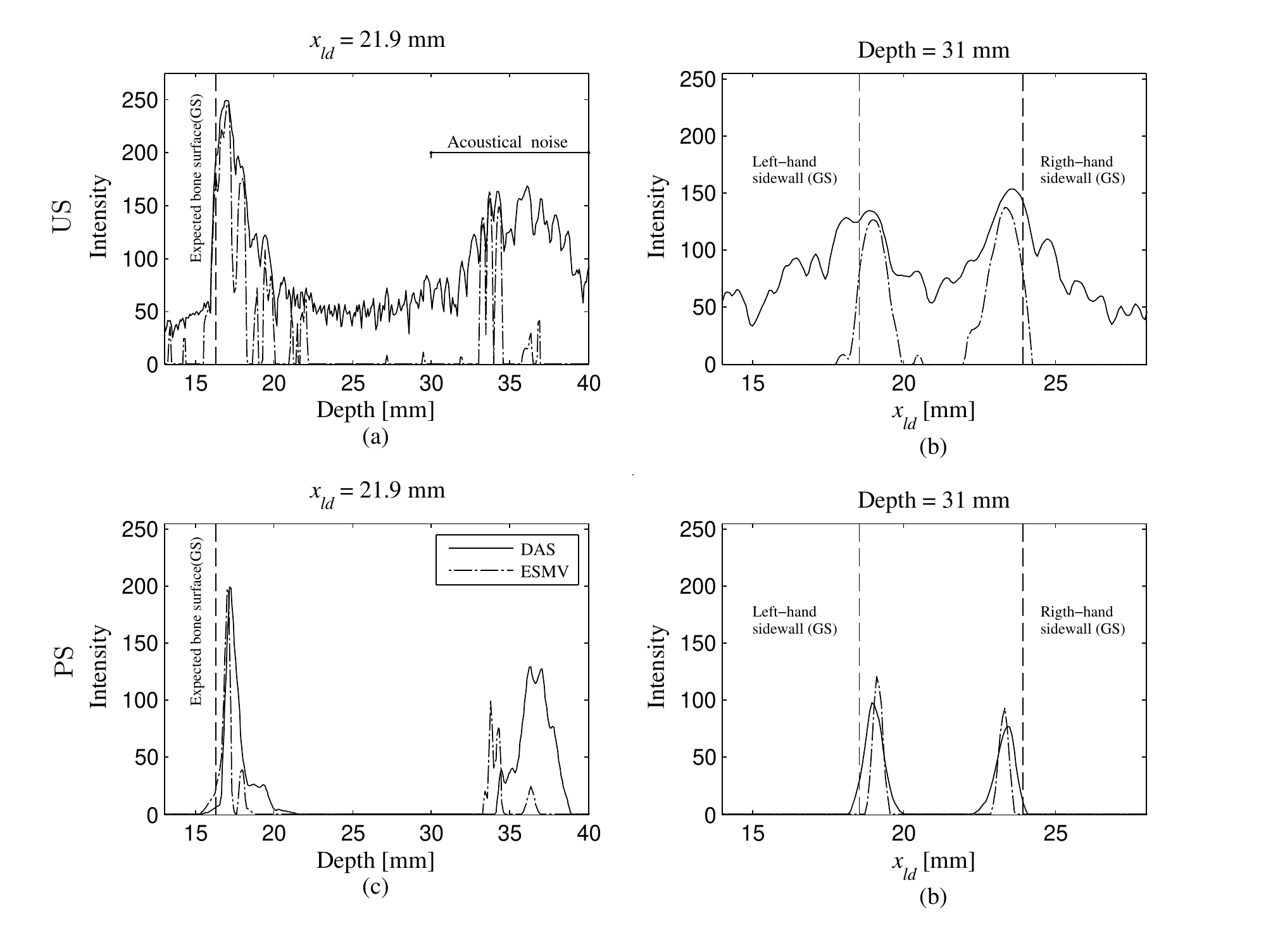}
\caption{Comparison of image lines in Figure~\ref{fig5} for the DAS and ESMV images, (a)US-images $x_{ld}$~=~14.32 mm, (b)US-images $x_{ld}$~=~21.80 mm, (c)PS-images depth~=~32.10 mm, and (d)PS-images depth~=~36.60 mm.}
\label{fig9}
\end{figure*}

 Fig.~\ref{fig9} presents two different image lines of the US  and PS  images in Fig.~\ref{fig5}. These lines are marked in Fig.~\ref{fig5}(c). In Fig.~\ref{fig9}(a), the mean  noise level,  measured between a depth of 30 mm  and 40 mm, is decreased from  69.01 in DAS to  14.33 in ESMV. Fig.~\ref{fig9}(b) shows a horizontal image line  at  $\rm Depth$~=~31 mm. In this  figure, the peaks at  $x_{ld}$~$\approx$~19.1 mm  and  $x_{ld}$~$\approx$~23.5 mm  indicate the  left-hand and  right-hand sidewalls at the corresponding depth. The signal sensitivity  at the  right-hand and left-hand  sidewalls are 126 and 136 for ESMV,  and 55 and 81 for DAS. In  Fig.~\ref{fig9}(c), the profile width measured at an intensity value of 150 is 0.59 mm  for DAS  and 0.29 mm for ESMV. The mean noise levels measured between  a depth of 17.45 mm  and 18 mm  are 20.35 and 4.62 for DAS and ESMV. In  Fig.~\ref{fig8}(d), the profile width at an intensity level of 50 is 0.71 m for DAS and 0.45 mm  for ESMV  for the left-hand sidewall.

\subsection{In-vivo images:}
\begin{figure*}[!htb]
\centering
\includegraphics[width=3.5in]{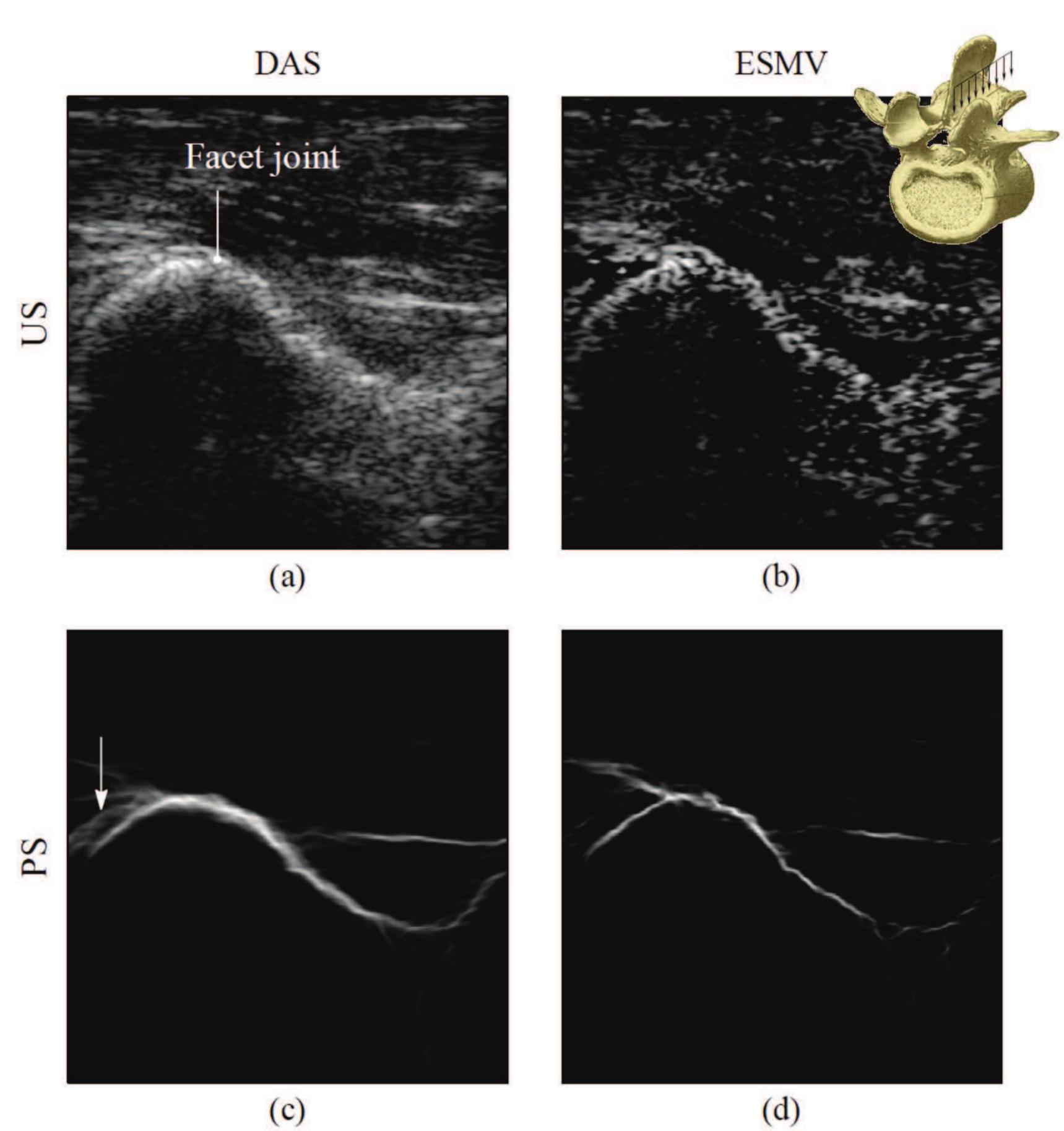}
\caption{Ultrasound and PS post-processed images of lamina. (a) DAS (US), (b) ESMV (US, $\beta=0.001\%$), (c) DAS (PS), and (d) ESMV (PS, $\beta=0.001\%$). The dynamic range is 50 dB and $L=M/2, K=2$ are used for (a) and (b).}
\label{fig10}
\end{figure*}

\begin{figure*}[!htb]
\centering
\includegraphics[width=3.5in]{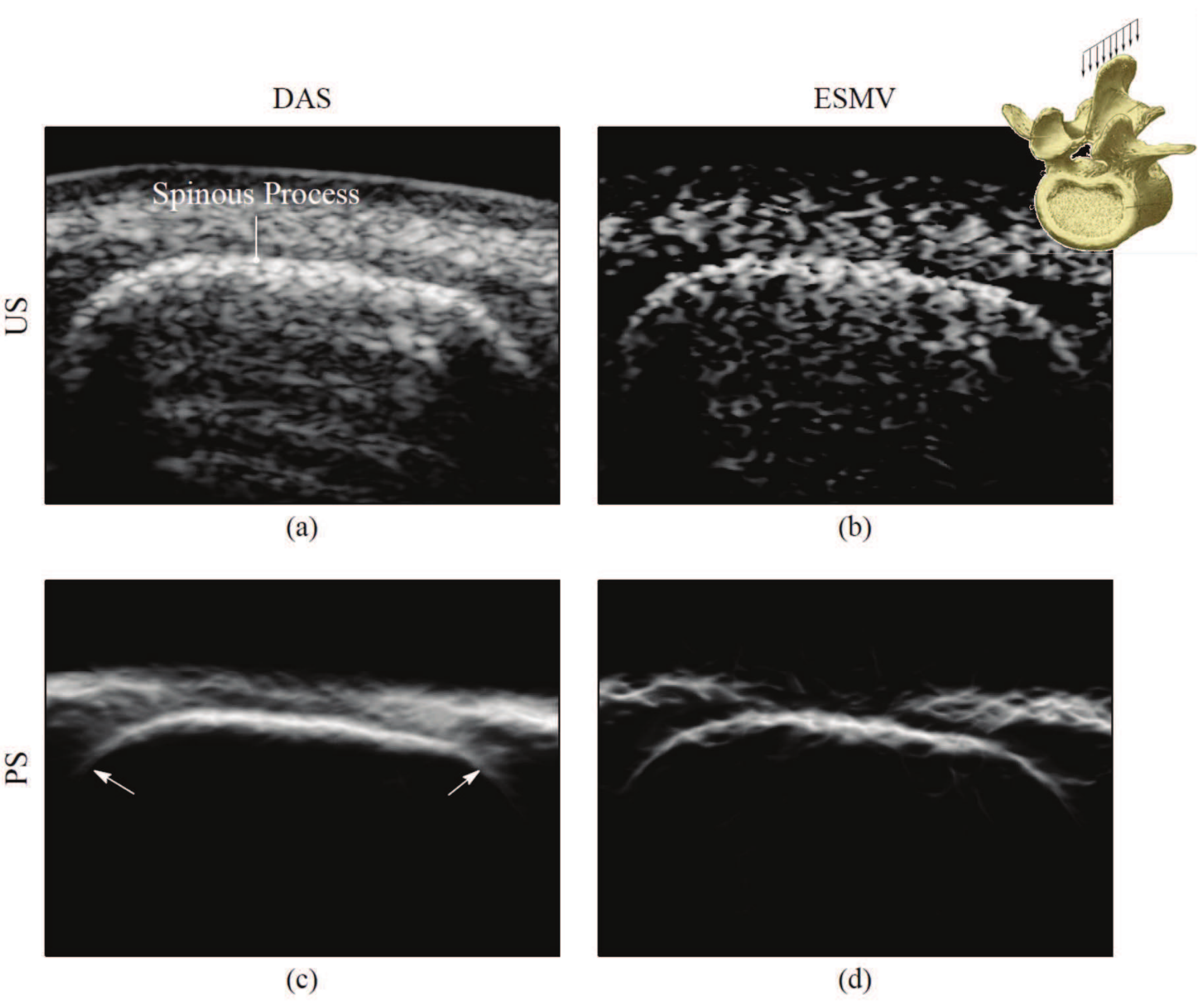}
\caption{Ultrasound and PS post-processed images of spinous process in sagittal direction for DAS and ESMV beamformers. (a) DAS (US), (b) ESMV (US, $\beta=0.001\%$), (c) DAS (PS), and (d) ESMV (PS, $\beta=0.001\%$). The dynamic range is 40 dB and $L=M/2, K=2$ are used for (a) and (b).}
\label{fig11}
\end{figure*}

\begin{figure*}[!htb]
\centering
\includegraphics[width=3.5in]{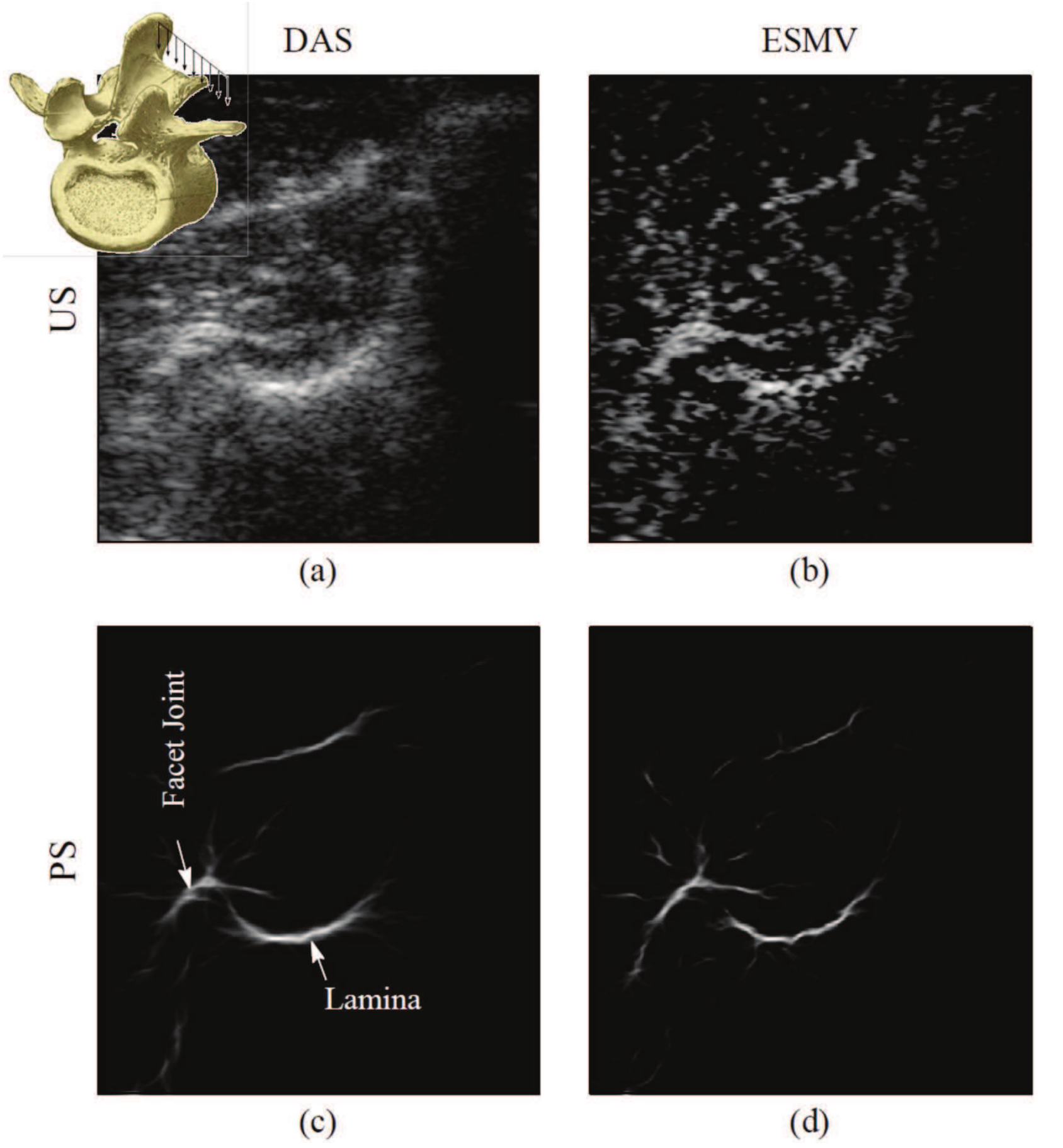}
\caption{Ultrasound and PS post-processed images of the spinous process in transversal direction. (a) DAS (US), (b) ESMV (US, $\beta=0.001\%$), (c) DAS (PS), and (d) ESMV (PS, $\beta=0.001\%$). The dynamic range is 50 dB and $L=M/2, K=2$ are used for (a) and (b).}
\label{fig12}
\end{figure*}

Figs.~\ref{fig10}~-~\ref{fig12} demonstrate a qualitative comparison between PS images obtained from DAS and ESMV beamformers. Fig.~\ref{fig10} shows images of a lamina in  the  sagittal direction. Fig.~\ref{fig10}(a) corresponds to the DAS image and Fig.~\ref{fig10}(b) demonstrates the ESMV image for $\beta=0.001\%$. It can be seen that in the ESMV image,the  amount of the speckle around the bone surface is reduced. Figs.~\ref{fig10}(c)~and~(d) show PS images obtained from Figs.~\ref{fig10}(a)~and~(b). It is observed that the ESMV beamformer improves the bone surface and results in a thinner definition of the bone boundary. Also on the left-hand side of the DAS image (marked with a white arrow) some unwanted features are observed, which have been removed in Fig.~\ref{fig10}(d).\\

Figs.~\ref{fig11} shows sagittal plane images of spinous process. Fig.~11(a) shows the DAS image.  Fig.~11(b) demonstrate the ESMV image with $\beta=0.001\%$. Comparing with Fig.~10(a), in this image the speckle around the bone surface is reduced while the structure of the bone is preserved. Fig.~11(c) shows the PS image obtained from the DAS image. In this image the bone surface is smeared out and the boundaries are not well delineated, whereas in Fig.~11(d) the bone surface is reasonably well isolated from the connective tissue on the top of the surface. In Fig.~10(c), the bone boundary, on both side of the spinous process marked with white arrows, is thick and unclear. In comparison, in Fig.~11(d) the bone boundary is sharper and a prolongation of the surface is observed. In a similar manner in Fig.~\ref{fig11}(d) the sharpness of the bone surface is increased for smaller $\beta$, and the surface is somewhat better isolated from the connective tissue in comparison with Fig.~\ref{fig11}(c).\

Fig.~\ref{fig12} shows an image of the lamina in the transversal direction. Comparing the US images, we observe a superior  isolation of  the bone  surface  from  surrounding  soft tissue in the ESMV image. Comparing the PS images, in  Fig.~\ref{fig12}(d),  we observe a delineation of  the  facet joint    on  left -hand side,  and the  lamina boundary on  the  right-hand side. Further,  in Fig.~\ref{fig12}(d),  an improved  isolation  between the facet joint   and   the lamina,   and a sharper  definition  of the bone boundaries  is  observed. \\

\section{Discussion}
\label{Discuss}
There is a potential for the ESMV beamformer to enhance the bone edges in  US images, but the performance of this beamformer depends on the signal subspace estimation. From Figs.~\ref{fig3}(b)~-~(d), we observe that by using a small threshold value the bone structure is preserved while the speckle in its neighborhood is reduced. This effect which has been discussed in~\cite{Mehdizadeh:unpublished} can give rise to images with enhanced edges but distorted speckle patterns [Figs.~\ref{fig3}(a)~-~(d)].  A very small thresholding value results in a rank-1 signal subspace [Fig.~\ref{fig3}(d)], {i.e.} just the largest eigenvalue is used for the signal subspace estimation in (\ref{eq:eq7}). Thus, since detection of edges is the main purpose, regardless of the speckle pattern, a rank-1 signal subspace can enhance the bone edges images obtained from the ESMV beamformer [Figs.~\ref{fig10}(b)~-~\ref{fig12}(b)]. This is beneficial for  post-processing techniques, {e.g.}~the phase symmetry method, for extracting or locating the bone surfaces.\\
In the simulated images, Fig.~\ref{fig3}, because of the specular  reflection, some parts of the  vertebra sidewalls are missed. In Fig.~\ref{fig3}(a), the  coherent scattering from  the perpendicular surfaces   to  the beams  result in  echoes  with  the  higher intensities, e.g., the  the  spinous process top,  and parts of  the lamina  located at $x_{ld} ={\pm}10$~mm. In  this  simulation setup, for  each  triangle,   the scatterers are  equally spaced  and located in-plane, and all  have equal  scattering strengths. That is, roughness effects are not considered in these images. However,  the angle between  the triangular surface elements can partly introduce the roughness to our simulation model.\\    
From Fig.~\ref{fig3}(h), and Figs.~\ref{fig10}(d)~-~\ref{fig12}(d) we observe that the bone surfaces which are extracted from the ESMV are sharper, the bone boundaries are thinner, and they are reasonably well isolated from the connective tissue in comparison  with  the DAS one. Also, this setup shows more details of the vertebra geometry, e.g. in Figs.~\ref{fig4}(d)~-~\ref{fig5}(d) the spinous process  geometry  is  well preserved ( at top  of the images) .\\
The registration with CT-contours, shown in Figs.~\ref{fig4}~and~\ref{fig5}, suggests that the ultrasound bone response appears within the CT-contours. The PS filtered bone surface is delineated at the maximum of the ultrasound bone response, which places it even further inside the CT-surface. This behavior of the PS-filter is expected from its mathematical formulation as it identifies the maximum of the response in the signal rather than its rising side. The other reason for the observed bias is due to the registration error as pinpointing the balls' tip accurately in the US images was  more  difficult than in the CT dataset.
In this study  the PS  parameters  are assigned  empirically,  and  finding an optimal  setup  for the log-Gabor  filter may be a challenge. In~ \cite{Hacihaliloglu:2011b}, an  automated procedure for selecting  the filter parameters has been investigated, which  can    ease the  filter tuning procedure.\\ 
 From  Fig.~\ref{fig3}, Fig.~\ref{fig4}, and Table~\ref{tab:TABEL0},  we observe  that the surface localization error in  both  DAS  and   ESMV  are in the  same order  of magnitude, but  narrower  distribution of the intensity values in region A  indicates  shaper  boundaries in the ESMV images. Further, the noise  level is much lower  inside / outside of the bone in the ESMV  images.\\  
The post-processed images in Figs.~\ref{fig10}~-~\ref{fig12}, demonstrate that there is a potential for the phase symmetry technique to reasonably well exploit the spinal structure from US images. This can result in an enhanced 3D reconstruction of the spinal anatomy, which facilitates level detection procedure in minimally invasive spinal surgeries~\cite{Otake:2012}, and registration of preoperative CT or MR images to intraoperative US in neuro-navigation surgeries~\cite{Winter:2008}. Furthermore, the superior separation of the bone surface from the connective tissues achieved in Fig.~\ref{fig11}(d), can ease the model-based automated segmentation of the spine anatomy.\\
The use of direction-dependent thresholding as designed in ~\cite{Hacihaliloglu:2009}, was not implemented as preservation of minute anatomical structures was considered more important than further noise removal. Further, the automatic adaptive parameterization suggested in~\cite{Hacihaliloglu:2011b} was not tested for this work. Manual tuning of the different parameters provided satisfying results. The automated approach will be implemented in our further work.
\section{Conclusions}
\label{Conclusions}
We have explored the potential of a rank-1 ESMV beamformer, together with the phase symmetry post-processing method to enhance the spinal anatomy in ultrasound images. The suggested beamformer is independent of the thresholding factor, and its complexity is in the same order as for minimum variance beamformer. This beamforming setup can locate the spinal structure reasonably well while reducing the speckle from the surrounding tissue. Therefore, the phase symmetry filtering of these images can result in an improved definition of the boundaries and enhanced separation of the spinal anatomy from the neighboring connective tissues in comparison with the DAS technique. This shows that beamforming which is optimized for good visual appearance is not always optimal for feature extraction. This is therefore one of the first examples which demonstrates that it can be beneficial to do beamforming in a way which does not give the best visual appearance, but rather one that gives the best feature detection. If good optimization criteria can be defined, then future work could take this one step further by actually doing a joint optimization of the two operations in order to improve feature detection.\\


\begin{thebibliography}{1}


\bibitem{Amin:2003}
D.~V. Amin, T.~Kanade, A.~M. Digioia, and B.~Jaramaz, ``Ultrasound registration
  of the bone surface for surgical navigation,'' \emph{Computer Aided Surgery},
  vol.~8, no.~1, pp. 1--16, 2003.

\bibitem{Winter:2008}
S.~Winter, B.~Brendel, I.~Pechlivanis, K.~Schmieder, and C.~Igel,
  ``Registration of ct and intraoperative 3-d ultrasound images of the spine
  using evolutionary and gradient-based methods,'' \emph{IEEE Transactions on
  Evolutionary Computation}, vol.~12, no.~3, pp. 284--296, 2008.

\bibitem{Stewart:2009}
C.~Stewart Siu-Wa, ``Emergency bedside ultrasound for the diagnosis of rib
  fractures,'' \emph{The American journal of Emergency Medicine}, vol.~27,
  no.~5, pp. 617--620, 2009.

\bibitem{Eichenberger:2004}
U.~Eichenberger, M.~Greher, and M.~Curatolo, ``Ultrasound in interventional pain management",'' \emph{Techniques in Regional
  Anesthesia and Pain Management}, vol.~8, no.~4, pp. 171 -- 178, 2004.

\bibitem{Gofeld:2009}
M.~Gofeld, A.~Bhatia, S.~Abbas, S.~Ganapathy, and M.~Johnson, ``Development and
  validation of a new technique for ultrasound-guided stellate ganglion
  block,'' \emph{Regional Anesthesia and Pain Medicine}, vol.~34, no.~5, pp.
  475--479, 2009.

\bibitem{Bonsanto:2005}
M.~M. Bonsanto, R.~Metzner, A.~Aschoff, V.~Tronnier, S.~Kunze, and C.~R. Wirtz,
  ``3d ultrasound navigation in syrinx surgery - a feasibility study,''
  \emph{Acta Neurochirurgica}, vol. 147, no.~5, pp. 533--541, 2005.

\bibitem{Kolstad:2006}
F.~Kolstad, O.~M. Rygh, T.~Selbekk, G.~Unsgaard, and O.~P. Nygaard,
  ``Three-dimensional ultrasonography navigation in spinal cord tumor
  surgery,'' \emph{journal of Neurosurgery: Spine}, vol.~5, no.~3, pp.
  264--270, 2006.

\bibitem{Arzola:2007}
C.~Arzola, S.~Davies, A.~Rofaeel, and J.~C. Carvalho, ``Ultrasound using the
  transverse approach to the lumbar spine provides reliable landmarks for labor
  epidurals,'' \emph{Anesthesia and analgesia}, vol. 104, no.~5, pp. 1188--92,
  2007, comparative Study.

\bibitem{Tran:2009}
D.~Tran, K.-W. Hor, V.~A. Lessoway, A.~A. Kamani, and R.~N. Rohling, ``Adaptive
  ultrasound imaging of the lumbar spine for guidance of epidural anesthesia,''
  \emph{Computerized Medical Imaging and Graphics}, vol.~33, no.~8, pp.
  593--601, 2009.

\bibitem{Narouze:2010}
S.~N. Narouze, ``Ultrasound-guided interventional procedures in pain
  management: Evidence-based medicine,'' \emph{Regional Anesthesia and Pain
  Medicine}, vol.~35, no.~2, pp. 55--58, 2010.

\bibitem{Grau:2005}
T.~Grau, ``Ultrasonography in the current practice of regional anaesthesia,''
  \emph{Best Practice \& Research Clinical Anaesthesiology}, vol.~19, no.~2,
  pp. 175--200, 2005.

\bibitem{Otake:2012}
Y.~Otake, S.~Schafer, J.~W.~Stayman, W.~Zbijewski, G.~Kleinszig, A.~Graumann,
  A.~J. Khanna, and J.~H.~ Siewerdsen, ``Automatic localization of target vertebrae in
  spine surgery using fast ct-to-fluoroscopy (3d-2d) image registration,''
  \emph{SPIE Medical Imaging}, vol. 8316, 2012.

\bibitem{Mauldin:2012}
F.~Mauldin, K.~Owen, M.~Tiouririne, and J.~Hossack, ``The effects of transducer
  geometry on artifacts common to diagnostic bone imaging with conventional
  medical ultrasound,'' \emph{Ultrasonics, Ferroelectrics and Frequency
  Control, IEEE Transactions on}, vol.~59, no.~6, pp. 1101 --1114, june 2012.

\bibitem{Kowal:2007}
J.~Kowal, C.~Amstutz, F.~Langlotz, H.~Talib, and M.~G. Ballester, ``Automated
  bone contour detection in ultrasound b-mode images for minimally invasive
  registration in computer-assisted surgery-an in vitro evaluation,'' \emph{The
  International journal of Medical Robotics and Computer Assisted Surgery},
  vol.~3, no.~4, pp. 341--348, 2007.

\bibitem{Jain:2004}
A.~K. Jain and R.~H. Taylor, ``Understanding bone responses in b-mode
  ultrasound images and automatic bone surface extraction using a bayesian
  probabilistic framework,'' in \emph{Medical Imaging 2004: Ultrasonic Imaging
  and Signal Processing}, vol. 5373.\hskip 1em plus 0.5em minus 0.4em\relax San
  Diego, CA, USA: SPIE, 2004, pp. 131--142.

\bibitem{Xiao:2004}
Z.~Xiao and Z.~Hou, ``Phase based feature detector consistent with human visual
  system characteristics,'' \emph{Pattern Recognition Letters}, vol.~25,
  no.~10, pp. 1115 -- 1121, 2004.

\bibitem{Hacihaliloglu:2009}
I.~Hacihaliloglu, R.~Abugharbieh, A.~J. Hodgson, and R.~N. Rohling, ``Bone
  surface localization in ultrasound using image phase-based features,''
  \emph{Ultrasound in Medicine \& Biology}, vol.~35, no.~9, pp. 1475--1487,
  2009.

\bibitem{Synnevag:2009}
J.~F. Synnevag, A.~Austeng, and S.~Holm, ``Benefits of minimum-variance
  beamforming in medical ultrasound imaging,'' \emph{IEEE Transactions on
  Ultrasonics, Ferroelectrics and Frequency Control}, vol.~56, no.~9, pp.
  1868--1879, 2009.

\bibitem{Capon:1969}
J.~Capon, ``High-resolution frequency-wavenumber spectrum analysis,''
  \emph{proc. IEEE}, vol.~57, no.~8, pp. 1408--1418, 1969.

\bibitem{Wang:2005}
Z.~Wang, J.~Li, and R.~Wu, ``Time-delay- and time-reversal-based robust capon
  beamformers for ultrasound imaging,'' \emph{IEEE Transactions on Medical
  Imaging}, vol.~24, no.~10, pp. 1308--1322, 2005.

\bibitem{Vignon:2008}
F.~Vignon and M.~R. Burcher, ``Capon beamforming in medical ultrasound imaging
  with focused beams,'' \emph{IEEE Transactions on Ultrasonics, Ferroelectrics
  and Frequency Control}, vol.~55, no.~3, pp. 619--628, 2008.

\bibitem{Asl:2010}
B.~Mohammadzadeh~Asl and A.~Mahloojifar, ``Eigenspace-based minimum variance
  beamforming applied to medical ultrasound imaging,'' \emph{IEEE Transactions
  on Ultrasonics, Ferroelectrics and Frequency Control}, vol.~57, no.~11, pp.
  2381--2390, 2010.

\bibitem{Cheng:1997}
L.~Cheng-Chou and L.~Ju-Hong, ``Eigenspace-based adaptive array beamforming
  with robust capabilities,'' \emph{IEEE Transactions on Antennas and
  Propagation}, vol.~45, no.~12, pp. 1711--1716, 1997.

\bibitem{Feldman:1994}
D.~D. Feldman and L.~J. Griffiths, ``A projection approach for robust adaptive
  beamforming,'' \emph{IEEE Transactions on Signal Processing}, vol.~42, no.~4,
  pp. 867--876, 1994.

\bibitem{Mehdizadeh:2012a}
S.~Mehdizadeh, A.~Austeng, T.~Johansen, and S.~Holm, ``Minimum variance
  beamforming applied to ultrasound imaging with a partially shaded aperture,''
  \emph{IEEE Transactions on Ultrasonics, Ferroelectrics and Frequency
  Control}, vol.~59, no.~4, pp. 683--693, 2012.

\bibitem{Mehdizadeh:2011}
------, ``Performance of adaptive beamformers for ultrasound imaging of a
  partially shaded object,'' \emph{IEEE Ultrasonics Symposium}, 2011.

\bibitem{Mehdizadeh:unpublished}
------, ``Eigenspace based minimum variance beamforming applied to ultrasound
  imaging of acoustically hard tissues,'' \emph{IEEE Transactions on Medical
  Imaging}, vol.~31, no.~10, pp. 1912--1921, 2012.

\bibitem{Featherstone:1997}
W.~Featherstone, H.~J. Strangeways, M.~A. Zatman, and H.~Mewes, ``A novel
  method to improve the performance of capon's minimum variance estimator,'' in
  \emph{Antennas and Propagation, Tenth International Conference on (Conf.
  Publ. No. 436)}, vol.~1, 1997, pp. 322--325.

\bibitem{Chang:1992}
L.~Chang and C.~C. Yeh, ``Performance of dmi and eigenspace-based
  beamformers,'' \emph{IEEE Transactions on Antennas and Propagation}, vol.~40,
  no.~11, pp. 1336--1347, 1992.

\bibitem{VanTrees:2002}
H.~L. Van~Trees, \emph{Optimum Array Processing - Part IV, Detection,
  Estimation, and Modulation Theory}.\hskip 1em plus 0.5em minus 0.4em\relax
  New York: John Wiley \& Sons, 2002.

\bibitem{Hacihaliloglu:2011a}
I.~Hacihaliloglu, R.~Abugharbieh, A.~J. Hodgson, and R.~N. Rohling, ``Automatic
  adaptive parameterization in local phase feature-based bone segmentation in
  ultrasound,'' \emph{Ultrasound in Medicine \& Biology}, vol.~37, no.~10, pp.
  1689--1703, 2011.

\bibitem{Jensen:1996}
J.~A. Jensen, ``Field: A program for simulating ultrasound systems,''
  \emph{Medical \& Biological Engineering \& Computing}, vol.~34, pp. 351--353,
  1996.

\bibitem{Lorensen:1987}
W.~E. Lorensen and H.~E. Cline, ``Marching cubes: A high resolution 3d surface
  construction algorithm,'' \emph{SIGGRAPH Comput. Graph.}, vol.~21, no.~4, pp.
  163--169, aug 1987.

\bibitem{Wagner:1988}
R.~F. Wagner, M.~F. Insana, and S.~W. Smith, ``Fundamental correlation lengths
  of coherent speckle in medical ultrasonic images,'' \emph{IEEE Transactions
  on Ultrasonics, Ferroelectrics and Frequency Control}, vol.~35, no.~1, pp.
  34--44, 1988.

\bibitem{Fitzpatrick:2009}
J.~M. Fitzpatrick, ``Fiducial registration error and target registration error
  are uncorrelated,'' in \emph{Society of Photo-Optical Instrumentation
  Engineers (SPIE) Conference Series}, vol. 7261, feb 2009.

\bibitem{Hacihaliloglu:2011b}
I.~Hacihaliloglu, R.~Abugharbieh, A.~J. Hodgson, R.~N. Rohling, and P.~Guy,
  ``Automatic bone localization and fracture detection from volumetric
  ultrasound images using 3-d local phase features,'' \emph{Ultrasound in
  Medicine \& Biology}, vol.~38, no.~1, pp. 128--144, 2011.




\end{thebibliography}
\end{document}